\documentclass[twocolumn,aps,superscriptaddress]{revtex4-2}
\usepackage{amsmath}
\usepackage{amssymb}
\usepackage{graphicx}
\usepackage{color}
\usepackage{adjustbox}
\usepackage{lipsum}
\usepackage{bm}
\usepackage{times}
\usepackage{multirow}
\usepackage{soul}
\usepackage{fontenc}
\usepackage{setspace}
\newcommand{\bea}{\begin{eqnarray}}
\newcommand{\eea}{\end{eqnarray}}
\newcommand{\bse}{\begin{subequations}}
\newcommand{\ese}{\end{subequations}}

\begin{document}
\title{Magnetic frustration driven by conduction carrier blocking in Nd$_2$Co$_{0.85}$Si$_{2.88}$}

\author{Mily Kundu}
%\email{mily.kundu@saha.ac.in}
\affiliation{Condensed Matter Physics Division, Saha Institute of Nuclear Physics, 1/AF Bidhannagar, Kolkata 700064, India}

\author{Santanu Pakhira}
%\email{spakhira@ameslab.gov}
\affiliation{Condensed Matter Physics Division, Saha Institute of Nuclear Physics, 1/AF Bidhannagar, Kolkata 700064, India}
\affiliation{Ames Laboratory, Iowa State University, Ames, Iowa 50011, USA}

%\cortext[cor1]{spakhira@ameslab.gov}

\author{Renu Choudhary}
\affiliation{Ames Laboratory, Iowa State University, Ames, Iowa 50011, USA}

\author{Shuvankar Gupta}
%\email{mily.kundu@saha.ac.in}
\affiliation{Condensed Matter Physics Division, Saha Institute of Nuclear Physics, 1/AF Bidhannagar, Kolkata 700064, India}

\author{Sudip Chakraborty}
%\email{mily.kundu@saha.ac.in}
\affiliation{Condensed Matter Physics Division, Saha Institute of Nuclear Physics, 1/AF Bidhannagar, Kolkata 700064, India}

\author{N. Lakshminarasimhan}
\affiliation{Electro-organic and Materials Electrochemistry Division, CSIR-Central Electrochemical Research Institute, Karaikudi 630 003, India}
\affiliation{Academy of Scientific and Innovative Research (AcSIR), Ghaziabad 201 002, India}

\author{R. Ranganathan}
\affiliation{Condensed Matter Physics Division, Saha Institute of Nuclear Physics, 1/AF Bidhannagar, Kolkata 700064, India}

\author{Duane D. Johnson}
\affiliation{Ames Laboratory, Iowa State University, Ames, Iowa 50011, USA}
\affiliation{Department of Materials Science \& Engineering, Iowa State University, Ames, Iowa 50011}

\author{Chandan Mazumdar}
%\email{chandan.mazumdar@saha.ac.in}
\affiliation{Condensed Matter Physics Division, Saha Institute of Nuclear Physics, 1/AF Bidhannagar, Kolkata 700064, India}
%\cortext[cor1]{Corresponding author}

\date{\today}

\begin{abstract}

The intermetallic compound Nd$_2$Co$_{0.85}$Si$_{2.88}$ having a triangular lattice could be synthesized in single-phase only with defect crystal structure. Investigation through different experimental techniques indicate the presence of two magnetic transitions in the system. As verified experimentally and theoretically, the high-temperature transition $T_{\rm H} \sim$ 140 K is associated with the development of ferromagnetic interaction between itinerant Co moments, whereas the low-temperature transition at $T_{\rm L} \sim$ 6.5 K is due to the coupling among Nd-4$f$ and Co-3$d$ moments, which is antiferromagnetic in nature. Detailed studies of temperature-dependent dc magnetic susceptibility, field dependence of isothermal magnetization, non-equilibrium dynamical behavior, \textit{viz.} magnetic relaxation, aging effect, magnetic-memory effect, and temperature dependence of heat capacity, along with density functional theory (DFT) calculations, suggest that the ground state is magnetically frustrated spin-glass in nature, having competing magnetic interactions of equivalent energies. DFT results further reveal that the 3$d$/5$d$-conduction carriers are blocked in the system and act as a barrier for the 4$f$-4$f$ RKKY interactions, resulting in spin-frustration. Presence of vacancy defects in the crystal are also conducive to the spin-frustration. This is an unique mechanism of magnetic frustration, not emphasized so far in any of the ternary R$_2$TX$_3$ (R=rare-earth, T=transition elements and X=Si, Ge, In) type compounds. Due to the competing character of the itinerant 3$d$ and localized 4$f$ moments, the compound exhibits anomalous field dependence of magnetic coercivity.

\end{abstract}

\maketitle
\section{Introduction}
Studies on magnetically frustrated systems are of significant interest due to the intriguing and unusual physical properties vis-\`{a}-vis quantum spin-liquid behavior, spin-ice state formation, spin-glass behavior, and multiple magnetic transitions~\cite{mydosh1996disordered,greedan2001geometrically,ramirez2001geometrical,martinez1997magnetic,moessner1998low,moessner2006geometrical}. Very recently, magnetic frustration is also found to be responsible for producing and stabilizing novel skyrmionic topological spin texture in different frustrated magnets~\cite{lohani2019quantum,ukleev2021frustration,hu2017creation,stajic2019skyrmions}. Frustration in a magnetic system can arise in the presence of antiferromagnetic (AFM) interactions in triangular or tetrahedral types of crystal lattices or in the presence of simultaneous competing magnetic interactions~\cite{lacroix2011introduction,blundell2003magnetism}. Generally, the presence of any kind of disorder in such magnetically frustrated systems lead to a spin-glass state formation~\cite{blundell2003magnetism}. Besides different critical novel phenomena, magnetically frustrated glassy systems are also known to exhibit different interesting non-equilibrium properties, like aging and magnetic memory that make the systems useful for technological applications~\cite{khan2014memory,pakhira2016large}. Thus, the investigation of different physical properties of most of the novel magnetically frustrated systems always remain of interest.

R$_2$TX$_3$ (R=rare-earth, T=transition elements and X=Si, Ge, In) type of systems are one where lattice geometry is incompatible with AFM interaction. Most of these systems crystallize in hexagonal AlB$_2$-type structure, where the R atoms form edge-sharing triangular lattices which are hexagonally stacked along the $c$-axis~\cite{pakhira2016large,majumdar2001observation,kundu2021complex,zhi2013structures,pakhira2018role,pakhira2018observation}. Between two such hexagonal layers of R atoms, T and X elements are randomly distributed, causing variation in the local electronic environment among the R atoms. Additionally, the nearest and next-nearest-neighbour distances between the R atoms are comparable, and thus conducive to magnetic frustration due to the presence of competing magnetic interactions. With triangular lattice geometry, competing ferromagnetic (FM) and AFM interactions and crystallographic disorders, many of the members of this series exhibit magnetically frustrated glassy behavior and manifest fascinating properties, like Kondo behavior, spin-glass behavior, multiple magnetic transitions, mixed-valence behavior, magnetic memory effect, and bidirectional frequency dependence of dynamical susceptibility~\cite{tien1997mass,zhang2017magnetism,mo2015magnetic,li2001magnetic,pakhira2017magnetic,pakhira2018unusual,pakhira2018role,pakhira2020resistivity,pakhira2019spatially}. Recently, novel skyrmionic phase was also observed in frustrated Gd$_2$PdSi$_3$~\cite{kurumaji2019skyrmion}, followed by a considerable number of fascinating studies on the same material \cite{hirschberger2020high,hirschberger2020topological,nomoto2020formation,zhang2020anomalous}.

As such, the study for other new isostructural compounds in order to search for different exotic physical properties is warranted. Notably, in all the studied R$_2$TX$_3$ systems (T = Ni, Pd, Pt, Cu, Rh) magnetism dominantly comes from localized 4$f$-states of rare-earth elements, while the contribution from the transition-element moment is rather negligible. In contrast, very recent study on Pr$_2$Co$_{0.86}$Si$_{2.88}$ \cite{kundu2021complex} reveals that itinerant moment associated with Co also considerably contribute to the magnetism of the compound along with the localized rare-earth moments and both these interactions compete with each other at low temperatures. The competing character of 3$d$ (due to Co) and 4$f$ (due to Pr) moments triggers Pr$_2$Co$_{0.86}$Si$_{2.88}$ to behave differently in comparison to other transition element based systems in this series. Moreover, this compound crystallizes in U$_2$RuSi$_3$-type structure, which has two non-equivalent R-sites, `1$a$' and `3$f$'. The non-identical environment around these two rare-earth sites leads to bond-disorder that may also contribute to competing magnetic-exchange interaction. In this context, it will be quite interesting to study other Co-based isostructural compounds to look for unusual physical properties associated with any novel magnetic state formation.

Here, we report the successful synthesis of Nd$_2$Co$_{0.85}$Si$_{2.88}$ belonging to the R$_2$TX$_3$ series. The compound has been studied through X-ray diffraction (XRD), scanning electron microscopy (SEM), energy-dispersive x-ray spectroscopy (EDX), magnetic measurements versus temperature $T$, Magnetic field $H$, and time $t$, with a special emphasis on the non-equilibrium dynamics, and heat capacity. For analysis, density functional theory (DFT) was used to calculate energy per cell and site magnetization for each spin configuration (with and without defects), vacancy formation energy, as well as spin-polarized local density of states. Our experimental studies, and theoretical analysis, indicate the simultaneous presence of competing magnetic interaction between 3$d$ and 4$f$ moments, resulting a magnetically frustrated glassy behavior at low temperatures and exhibits fascinating non-equilibrium dynamical behavior typically observed in spin-glass systems. The presence of both the magnetic contribution from localized Nd-moments and itinerant Co-moments results an unusual magnetic-field dependence of coercive field.

\section{Experimental Details}
Polycrystalline Nd-based and La-based (non-magnetic reference) samples were prepared using conventional arc-melting procedure, by taking appropriate amounts of the pure ($\geq$ 99.9 \%) constituent elements. To get a homogeneous formation of the samples, the ingots were remelted several times by flipping them after each melt. Then the samples were annealed in an evacuated quartz tube at 1000$^{\circ}$C for 10 days followed by normal cooling. XRD measurements were performed on a TTRAX-III rotating anode diffractometer (M/s. Rigaku Corp., Japan) using 9 kW power in the temperature range 15-300 K. Structural characterizations were performed using FullProf proggramme package \cite{rodriguez1993recent}, through the Rietveld analysis of the XRD patterns. The homogeneity and elemental composition of the prepared samples were investigated through scanning electron microscopy (SEM) in the instrument EVO 18 (M/s. Carl Zeiss, AG, Germany) attached with a energy dispersive x-ray spectroscopy (EDX) set-up (M/s. EDAX, USA). DC magnetic measurements have been carried out for different applied magnetic fields (- 70 kOe $\le$ magnetic field ($H$) $\le$ 70 kOe) in the temperature range 2-300 K, using an Evercool II VSM (M/s. Quantum Design Inc., USA) and SQUID VSM (M/s. Quantum Design Inc., USA). Conventional zero-field-cooled (ZFC) and field-cooled (FC) protocols were used for collecting the magnetization data of our system. Magnetic isotherms ($M$ $vs$. $H$) were taken in ZFC condition. Zero-field heat capacity data has been taken using a commercial Physical Property Measurement System (PPMS) (M/s. Dynacool, Quantum Design Inc., USA).

\section{\label{Sec:DFT} Computational Details}

To investigate the nature of the magnetic ground state and associated magnetic interactions, we apply density-functional theory (DFT) to the cell of Nd$_4$Co$_2$Si$_6$ (Fig.~\ref{structure_paper}), in various magnetic configurations. In addition, we address in this system the effect of limited vacancy formation and its effect on relative energies and magnetic stability by calculating the divacancy energy for the 2$\times$2$\times$1 supercell Nd$_{16}$Co$_7$Si$_{23}$, where we removed 1/8 Co (12.50\%) and 1/24 Si (4.167\%) – an effective composition of Nd$_2$Co$_{0.875}$Si$_{2.88}$, similar to that found experimentally. The DFT calculations are based on the projector-augmented wave (PAW) method \cite{kresse1999ultrasoft}, as implemented in the Vienna Ab-initio Simulation Package (VASP). For the exchange and correlation functional, the Perdew-Burke-Ernzerhof generalized gradient approximation (PBE-GGA) is used \cite{kresse1996efficient}. We also used spin-orbit coupling (SOC), an onsite electron correlation parameter, \textit{i.e.}, an effective Hubbard parameter ($\lvert$U-J$_{\rm ex} \rvert$ = 5 eV) for the strongly correlated Nd-4$f$ states. The present low-temperature experimental lattice parameters of $a=b= 8.081$ \AA~and $c=4.174$ \AA~are used in the DFT calculations. The convergence criterion for the self-consistent calculations is 10$^{-7}$ eV for the total-energy per cell, and an energy cutoff of 520 eV is used for the electronic wave functions. The $k$-point integration was performed using a tetrahedron method with Bloch corrections. A $\Gamma$-centered grid of 12$\times$12$\times$24 $k$-points was used for Brillouin zone sampling.

\section{Results and Discussions}
\subsection{\label{Sec:structure} Structural details and phase analysis}

\begin{figure}
 \centering
 \includegraphics[width = 3.2in]{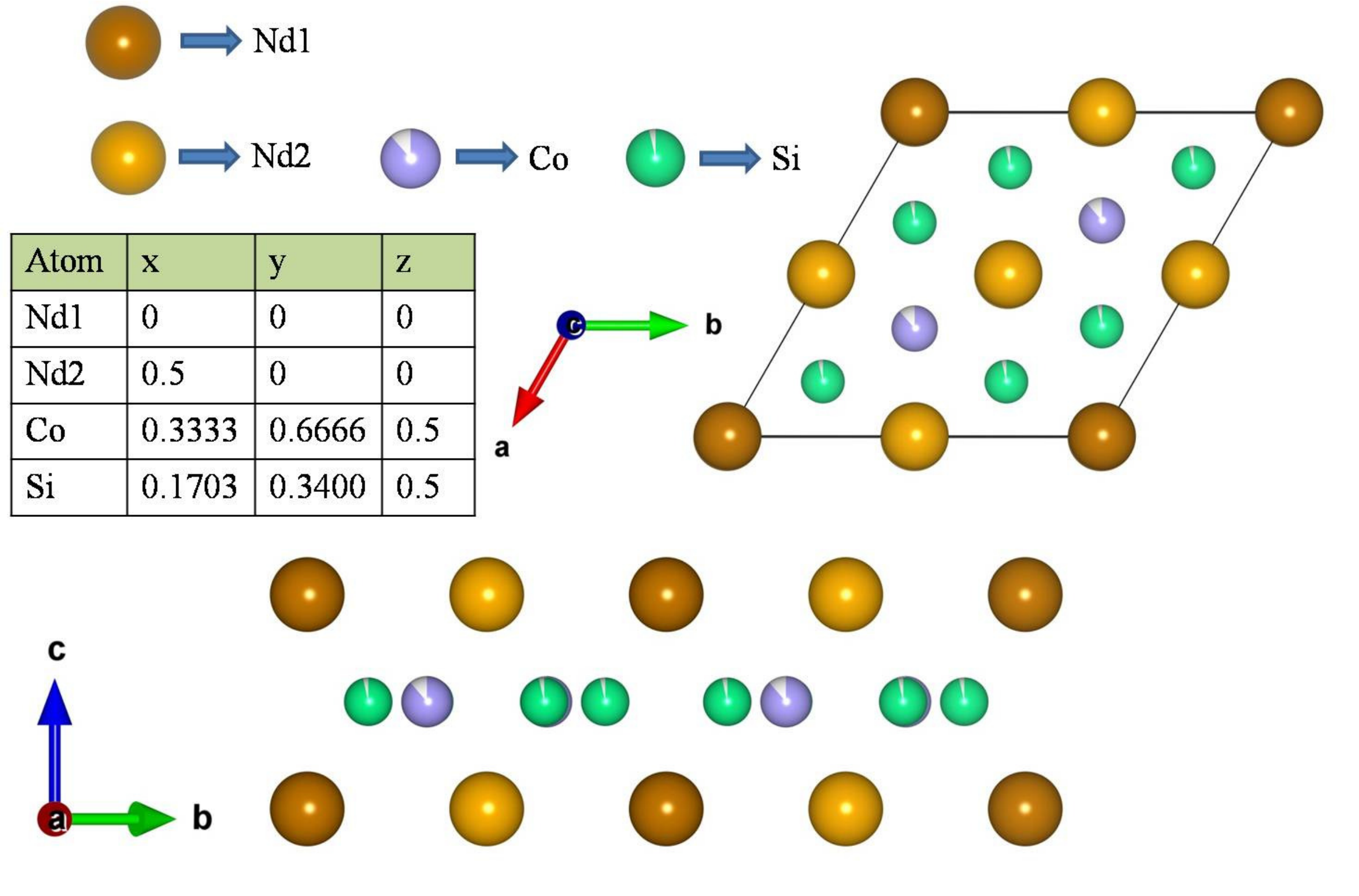}
 \caption{Crystal structure of Nd$_2$Co$_{0.85}$Si$_{2.88}$ belonging to the U$_2$RuSi$_3$-type structure (space group: \textit{P6/mmm}) which is a superstructure of AlB$_2$-type structure.}\label{structure_paper}
\end{figure}

\begin{figure}
 \centering
 \includegraphics[width = 3.3in]{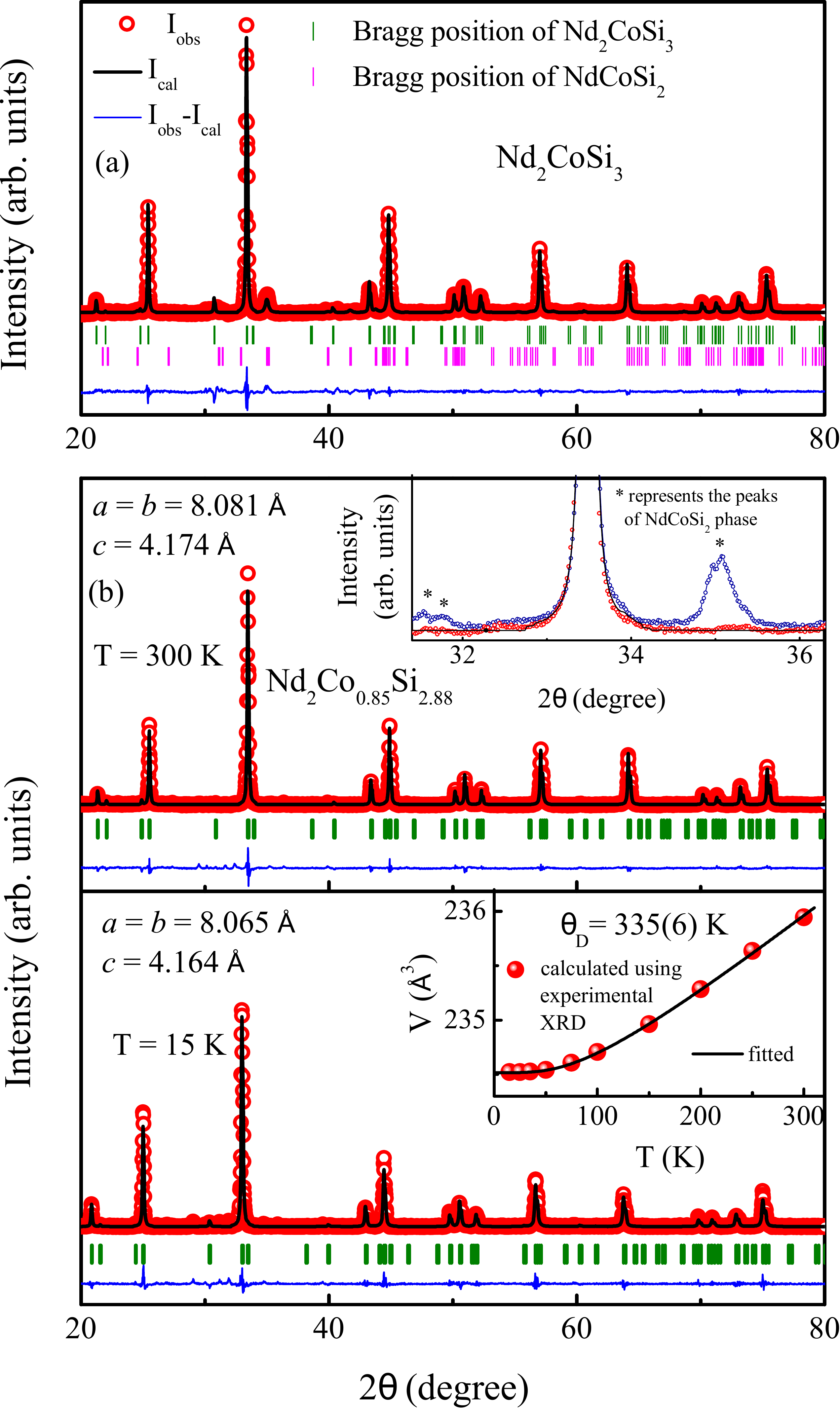}
 \caption{(a) Room temperature XRD pattern of Nd$_2$CoSi$_3$. (b) The XRD patterns of Nd$_2$Co$_{0.85}$Si$_{2.88}$ at $T$ = 300 K (top) and $T$ = 15 K (bottom) along with full-Rietveld analysis. Top inset shows the major peaks associated with the secondary phase NdCoSi$_2$ that is absent in defect formation. Bottom inset presents the temperature dependence of unit cell volume along with fit using Eq. \ref{volume}. Estimated errors are smaller than the symbol size.}\label{Nd_XRD}
\end{figure}

\begin{table}
	
	\caption{Crystallographic parameters and composition from XRD and SEM-EDX analysis at room-temperature.}
	\begin{center}
		\renewcommand{\arraystretch}{1.25}
		%\begin{ruledtabular}
		
		\begin{tabular}{l l}
			\hline\hline
			Compound &\hspace{0.6cm} Nd$_2$Co$_{0.85}$Si$_{2.88}$\\
			Structure &\hspace{0.6cm} U$_2$RuSi$_3$ type\\
			Space group &\hspace{0.6cm} \textit{P6/mmm}\\
			Lattice parameters & \\
			$a~({\rm \AA})$ &\hspace{0.6cm} 8.081 \\
			$c~({\rm \AA})$ &\hspace{0.6cm} 4.174 \\
			$R_{\rm f}$(\%) &\hspace{0.6cm} 2.29\\
			$R_{\rm {Bragg}}$(\%) &\hspace{0.6cm} 2.02\\
            $\chi^2$ &\hspace{0.6cm} 8.4\\
			\hline
			
		\end{tabular}
		%}
		%\begin{table*}[ht]
		%\centering
		\renewcommand{\arraystretch}{1.5}
		%\begin{ruledtabular}
		\resizebox{0.55\columnwidth}{!}{%
			\begin{tabular}{c|c}
				\hline
				\multicolumn{2}{c}{Estimated composition} \\
				\cline{1-2}
				Full-Rietveld & EDX\\
				\hline
				Nd$_2$Co$_{0.89}$Si$_{2.89}$  & Nd$_2$Co$_{0.87}$Si$_{2.87}$ \\
				\hline\hline
				
			\end{tabular}
		}
		%\end{ruledtabular}
	\end{center}
	\label{XRD_parameter}
	
\end{table}

\begin{figure}
 \centering
 \includegraphics[width = 3.3in]{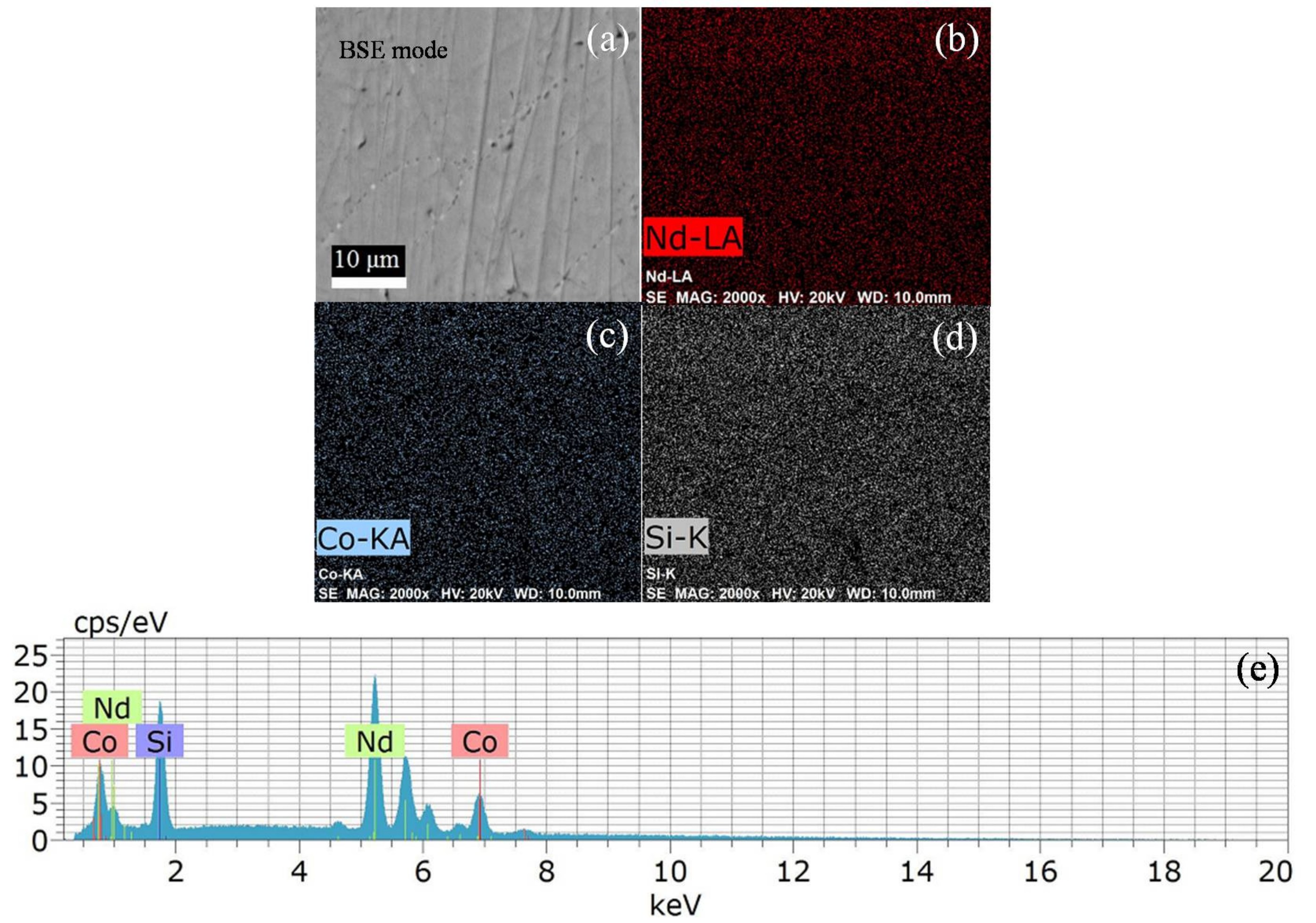}
 \caption{(a) Backscattered electron (BSE) image and (b-d) presents elemental mapping by SEM equipped with EDX on the polished surface of Nd$_2$Co$_{0.85}$Si$_{2.88}$ from the region presented in (a). (e) presents EDX results of Nd$_2$Co$_{0.85}$Si$_{2.88}$.}\label{Nd_EDX}
\end{figure}

Full Rietveld analysis of the room-temperature XRD pattern of stoichiometric composition Nd$_2$CoSi$_3$ indicates the presence of a few additional peaks those can not be indexed with the fully ordered U$_2$RuSi$_3$-type crystal structure ($a\approx 2c$), shown in Fig.~\ref{structure_paper}. Those additional peaks appear due to the presence of NdCoSi$_2$ (space group \textit{Cmcm}) secondary phase in addition to the main phase Nd$_2$CoSi$_3$ (U$_2$RuSi$_3$-type structure with space group \textit{P6/mmm}), as shown in Fig.~\ref{Nd_XRD}(a). The single phase nature of the compound could be achieved by deliberately introducing appropriate vacancies in the Co and Si sites (estimated by subtracting the secondary phase fraction in the stoichiometric material described above), \textit{i.e.}, forming a defect structure having nominal composition Nd$_2$Co$_{0.85}$Si$_{2.88}$. Single-phase character and homogeneity of the synthesized material was further confirmed by taking SEM image in back-scattered electron (BSE) mode and elemental mapping through EDX measurements. The elemental compositions obtained from full Rietveld analysis and EDX analysis are Nd$_2$Co$_{0.89}$Si$_{2.89}$ and Nd$_2$Co$_{0.87}$Si$_{2.87}$, respectively and are quite close to the starting elemental composition (Table~\ref{XRD_parameter}). It could be seen from the Fig.~\ref{structure_paper} that the nearest-neighbour and next-nearest-neighbour distances between the Nd-atoms are comparable along $a$ and $c$ directions. In addition to this, two non-identical Wyckoff sites of Nd atom have different environment that results in bond-disorder in the system. The crystal structure remains down to 15 K (Fig.~\ref{Nd_XRD}(b)), similar to many other members of the R$_2$TSi$_3$-type systems \cite{pakhira2016large,pakhira2018unusual}. The temperature dependence of lattice volume could be utilized to estimate the Debye temperature ($\theta_{\rm D}$) as the lattice volume of the system can be expressed as

\begin{align}\label{volume}
   V(T)&= \frac{\gamma U(T)}{K_{0}}+V_{0},
\end{align}

\noindent where  $V_0$ denotes the unit-cell volume at $T$= 0 K, $K_{0}$ is the bulk modulus, $\gamma$ signifies the Gr$\ddot{\rm{u}}$neisen parameter, and \textit{U(T)} is the internal energy that could be written as
\begin{eqnarray}
% \nonumber % Remove numbering (before each equation)
  U(T)&=&9nRT{\left(\frac{T}{\theta_{\rm D}}\right)}^3\int_{0}^{\theta_{D}/T}\frac{x^3}{e^x-1} dx\label{lattice_energy}
\end{eqnarray}

\noindent Here \textit{n} indicates the number of atoms per formula unit (f.u.). The fitting yields $\theta_{\rm D}$ to be 335(6) K, which lies in the same range reported for many R$_2$TSi$_3$-type of compounds~\cite{pakhira2016large,kundu2021complex,pakhira2018unusual}.

\subsection{\label{Sec:magnetization} DC Magnetic Susceptibility}
\begin{figure}[h!]
\centering
\includegraphics[width = 3.3in]{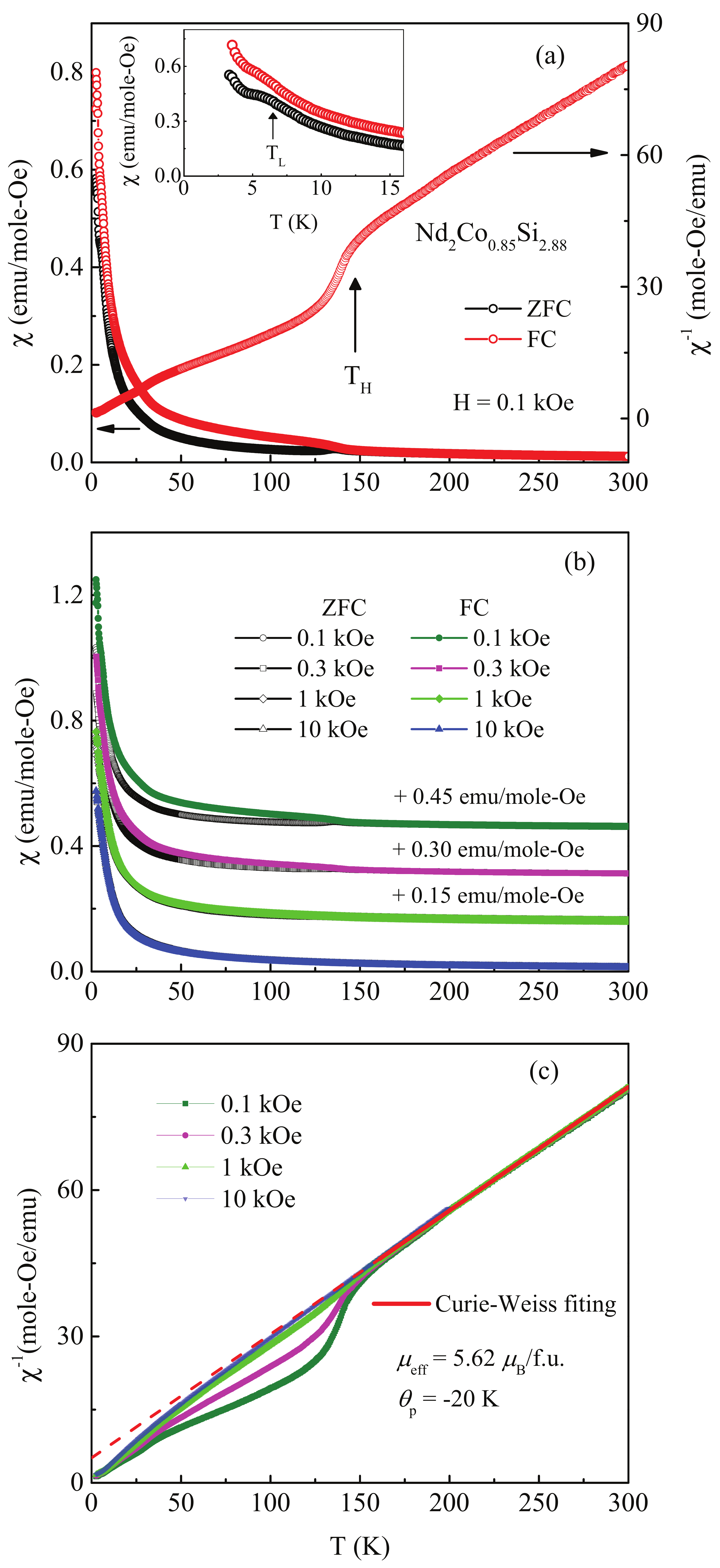}
\caption {(a) Left panel shows the temperature dependence of susceptibility ($\chi$) in ZFC and FC protocol at $H=0.1$~kOe and right panel represents the inverse susceptibility data for FC measurement. Inset shows an expanded view of $\chi (T)$ for both the ZFC and FC configuration at low temperature region. (b) $\chi(T)$ in ZFC and FC protocol for various applied magnetic fields. The data are separated from each other manually to have a clear view. (c) $\chi^{-1}(T)$ for FC mode at different magnetic fields. Curie-Weiss fitting of $\chi^{-1}(T)$ is presented for 160-300 K at $H = 1$~kOe.}\label{Nd_MT}
\end{figure}

The dc magnetic susceptibility ($\chi \equiv M/H$) of Nd$_2$Co$_{0.85}$Si$_{2.88}$ measured in zero-field-cooled (ZFC) and field-cooled (FC) protocols for $H=0.1$~kOe and other magnetic fields are shown in Fig.~\ref{Nd_MT}(a) and Fig.~\ref{Nd_MT}(b), respectively. The $\chi (T)$ behavior measured for $H = 0.1$~kOe indicates the presence of two cusp-like feature, one around $T_{\rm H} \sim$ 140~K and another at a lower temperature $T_{\rm L} \sim$ 6.5 K (Fig.~\ref{Nd_MT}(a) inset). The ZFC and FC magnetic susceptibility curves diverge from each other below $\sim$ 140 K. The bifurcation between $\chi_{\rm {ZFC}}$ and $\chi_{\rm {FC}}$, however, is not manifested when measured at $H \geq 1$ kOe. Interestingly, both the $\chi_{\rm {ZFC}}$ and $\chi_{\rm {FC}}$ increases with lowering temperature even below $T_{\rm L}$, indicating the anomaly at $T_{\rm L}$ to be either due to the formation of magnetically glassy state with short range correlation and/or having an additional lower temperature ordering similar to that reported earlier in Nd$_2$Ni$_{0.94}$Si$_{2.94}$ \cite{pakhira2018chemical}. As the material is essentially single phase, the temperature dependence of magnetic susceptibility behavior clearly suggests the presence of at least two different magnetic transitions in the system, one at $T_{\rm L}$ and another at $T_{\rm H}$. The anomaly found in the high-temperature region, $T_{\rm H} \sim$ 140~K, at $H = 0.1$~kOe is much more prominent in the inverse susceptibility behavior (Fig.~\ref{Nd_MT}(a)) that goes on diminishing with increasing strength of magnetic fields (Fig.~\ref{Nd_MT}(c)). Similar behavior was also observed in Pr$_2$Co$_{0.86}$Si$_{2.88}$, where it was established to originate due to the development of short-range interaction among itinerant Co-3$d$ moments \cite{kundu2021complex}. It is quite likely that the magnetic transition occurring around $T_{\rm H}$ in Nd$_2$Co$_{0.85}$Si$_{2.88}$ might also have the very similar origin. To confirm the presence of Co moment, the temperature dependence of inverse susceptibility is fitted with a Curie-Weiss (CW) law
\bea
\chi(T) = \frac{C}{T-\theta_{\rm p}},
\label{Eq.ModCurieWeiss}
\eea
where $\theta_{\rm p}$ is the Weiss temperature, and $C$ is Curie constant given by
\bea
C&=&\frac{N_{\rm A} {g}^2S(S+1)\mu^2_{\rm B}}{3k_{\rm B}} = \frac{N_{\rm A}\mu^2_{\rm {eff}}\mu^2_{\rm B}}{3k_{\rm B}},
\label{Eq.Cvalue1}
\eea

\noindent where $N_{\rm A}$ is Avogadro's number, $g$ is the spectroscopic splitting factor, $S$ is the spin quantum number, $k_{\rm B}$ is Boltzmann's constant,  and $\mu_{\rm eff}$ is the effective moment of a spin in units of Bohr magnetons, $\mu\rm_B$. The linear CW fitting (Fig.~\ref{Nd_MT}(c)) in the temperature range 160-300 K yields, $\theta_{\rm p}= -20.0(5)$ K and $\mu_{\rm eff}= 5.62(2)~\mu_{\rm B}$/f.u. Considering only rare-earth Nd ion to be the moment carrying element, the effective moment per Nd-ion is found to be 3.97(2)~$\mu_{\rm B}$/Nd-ion, which is slightly higher than the theoretical value of 3.62~$\mu_{\rm B}$, expected for one free Nd$^{3+}$ ion. The 3$d$-spins from Co-atoms could be the likely source of the additional $0.35~\mu_{\rm B}$, estimated from the analysis of magnetic susceptibility data.

Inserting the Gaussian (cgs) values of the fundamental constants (where ${\rm 1~emu = 1~G~cm^3/mol}$) into Eq.~(\ref{Eq.Cvalue1}), the Curie constant per mole of spins is expressed as
\bea
C {\rm (cm^3~K/mol)} \approx \frac{\mu^2_{\rm eff}}{8} ~(\mu^2_{\rm_ B}/{\rm f.u.}),
\label{Eq.Cvalue2}
\eea
and hence
\bea
\mu_{\rm eff}~ {\rm (\mu_B/f.u.)} \approx \sqrt{8C}
\label{Eq.mueff}
\eea

\noindent For spins~$S$ with $g=2$, the isotropic Curie constant in units of ${\rm cm^3\,K/mol\,spins}$ is
\bea
C = 0.5002 S(S+1).
\label{Eq.Curieconstant}
\eea

\noindent Thus, in a local moment system with minimum spin value $S=1/2$, one expect $C \approx 0.375~{\rm cm^3\,K/mol\,spins}$, which gives $\mu_{\rm eff} \approx 1.73~{\rm \mu_B/f.u.}$ The much lesser value of $\mu_{\rm eff} \sim 0.35~\mu_{\rm B}$ obtained per Co-spins in Nd$_2$Co$_{0.85}$Si$_{2.88}$ clearly suggests that the magnetic moment associated with Co is itinerant in nature. The nature and estimated value of Co moments is in good agreement with the theoretical calculations, discussed in sec.~\ref{Sec:DFT}.

Note, the estimated CW temperature ($\theta_{\rm p}$) is significantly smaller than the transition temperature $T_{\rm H}$. In this system, although both itinerant Co-moment and localized Nd-moment contribute in towards the magnetism, the effect of localized 4$f$-spins of Nd have a grossly overwhelming influence over the weak nature of itinerant 3$d$-spins of Co. The CW temperature of a magnetic system reflects the strength of exchange coupling/interaction present the system. In the present case, RKKY exchange interaction strength between the Nd-atoms is significantly stronger than the coupling between Co-spins. Hence, although a magnetic transition associated with Co-atoms is observed at $T_{\rm H} \sim 140$~K, the estimated CW temperature is significantly smaller. The negative sign of the CW temperature is in accordance with the antiferromagnetic coupling between the Nd and Co atoms, as inferred from our theoretical analysis.

The observed low-temperature transition at $T_{\rm L}$ is relatively broad in temperature than that generally observed for a long-range magnetic ordering and only persists at lower applied magnetic fields. At higher fields a field-induced behavior is observed. The susceptibility cusp at $T_{\rm L}$ and relatively fragile nature of the ground state spin configurations indicate a metastable state formation below $T_{\rm L}$ for the compound. Our DFT calculations (discussed in section~\ref{Sec:DFT}) further confirm that the magnetic ground state of the compound is indeed magnetically frustrated in nature and consists of multiple equivalent low-energy configurations.

\begin{figure*}[ht!]
\centering
\includegraphics[width = 5.2in]{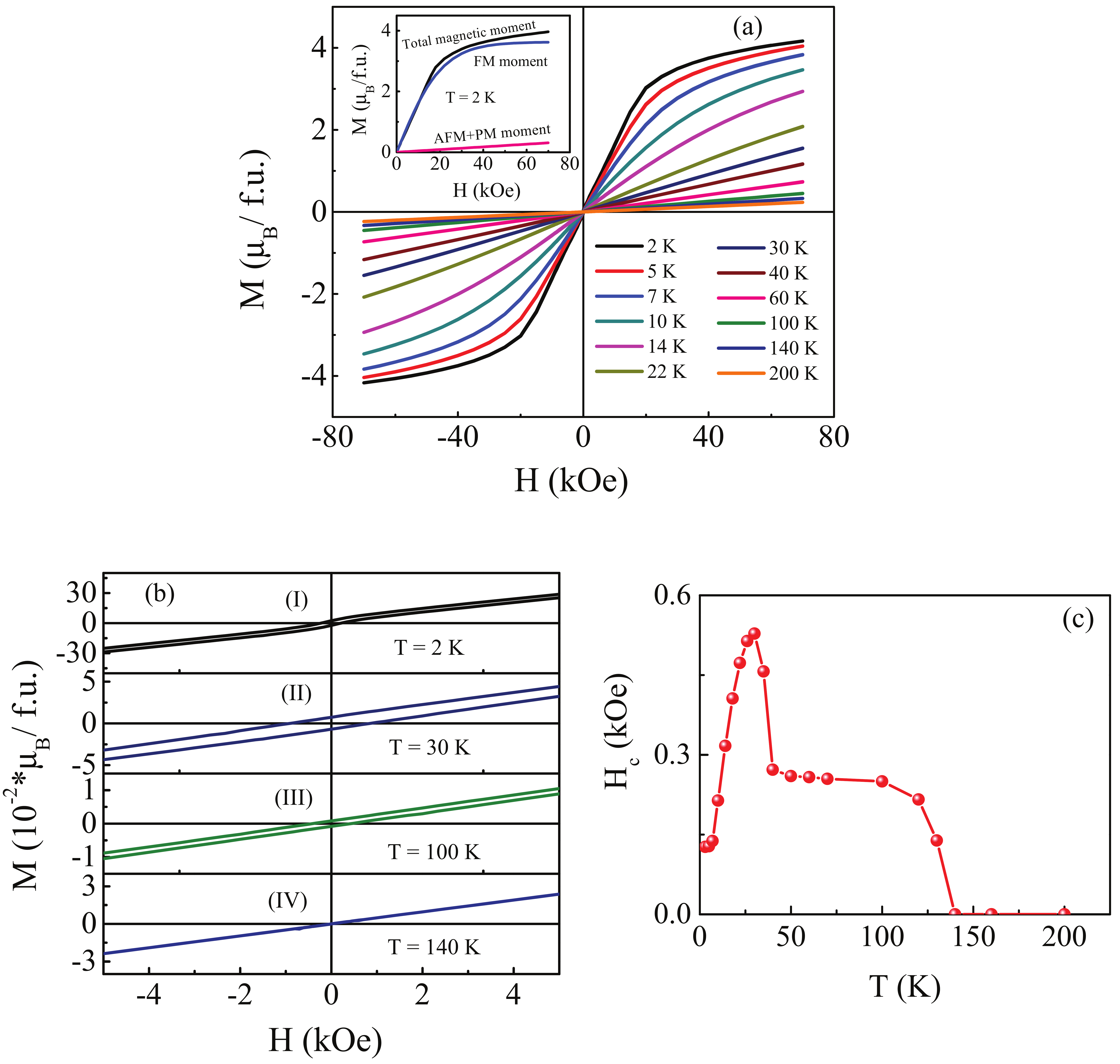}
\caption {(a) Magnetic field dependence of magnetization at different temperatures. Inset shows the isothermal magnetization curve at $T$ = 2 K along with the estimated FM and AFM/PM contributions. (b) Isothermal magnetization at some selected temperatures, $T$ = (I) 2 K; (II) 30 K; (III) 100 K and (IV) 140 K, close to the low field region. (c) Temperature dependence of coercive field. Solid line is guide to the eye.}\label{Nd_MH}
\end{figure*}

\subsection{Field dependence of isothermal magnetization}

To understand the field dependence of magnetic ground state further, the field variant magnetization is measured at different temperatures and are shown in Fig.~\ref{Nd_MH}(a). The $M(H)$ behavior for $T \leq T_{\rm L}$ is linear in the low field region and tends to saturate at higher fields. The overall nature of the magnetic isotherms at low temperatures are quite similar to those of earlier reported for different R$_2$TX$_3$-type of compounds having fragile frustrated glassy magnetic ground states and high-field-induced ferromagnetic behavior \cite{pakhira2016large,kundu2021complex}. The $M$-$H$ curves gradually turn out to be linear with increasing temperature of the system and linearity in the whole field range is only observed for $T \geq 140$~K. The field dependent isothermal magnetization at $T = 2$~K reaches a value of 4.17 $\mu_{\rm B}$/f.u. for 70 kOe applied field, that is itself much lower than the theoretical saturation moment value of 7.24 $\mu_{\rm B}$/f.u. expected for the system considering only Nd$^{3+}$ contributes to the magnetism. Such lower value suggests the critical role of crystalline electric field (CEF) in the system and/or an antiparallel coupling between localized Nd moment and itinerant Co moment. The progressive enhancement in $M(H)$ even at a strong magnetic field of 70 kOe further supports the presence of itinerant magnetism in the system \cite{takahashi2013spin}. As shown in the inset of Fig.~\ref{Nd_MH}(a), the low-temperature isotherm at 2 K, could be well described using a tangent-hyperbolic and a linear term that represents the ferromagnetic and antiferromagetic (or paramagnetic) interaction, respectively.

A careful investigation of the magnetic isotherms measured as a function of temperatures (Fig.~\ref{Nd_MH}(b)) reveal the presence of finite coercivity $H_{\rm c}$ in the system for $T \leq T_{\rm H}$. At the lowest measured temperature $T = 2$~K, $H_{\rm c}$ is found to be $\sim 125$~Oe. Interestingly, the $H_{\rm c}$ rapidly increases with increasing temperature, reaches a maximum at around 30 K followed by a sharp fall around 50 K and remains almost constant up to 100 K, and then gradually vanishes at $\sim$ 140 K (Fig.~\ref{Nd_MH}(c)). Such a behavior of coercivity is quite similar to that observed for isostructural Pr$_2$Co$_{0.86}$Si$_{2.88}$ except the fact that in the present case one sees finite coercivity in the entire temperature region $T\le$ $T_{\rm H}$, whereas in Pr-analogue $H_{\rm c}$ appears at a temperature much below $T_{\rm H}$ and also vanishes below $T_{\rm L}$ \cite{kundu2021complex}. The presence of $H_{\rm c}$ for $T \leq T_{\rm H}$ is associated with the gradual development of ferromagnetic correlation among the itinerant Co-moments. It is well known that for a ferromagnetic phase, atomic spins are aligned parallely within separate domains. Under the application of magnetic field, the spins of all the domains tend to align along the field direction. However, on removal of external magnetic field, the aligned spins in the domains cannot attain its previous magnetic state that results in appearance of coercivity. As the temperature of the system decreases, as expected, the ferromagnetic interaction firms up further between the Co-moments causing the enhancement of $H_{\rm c}$. Below around 100 K, $H_{\rm c}$ remains almost temperature independent down to $\sim$ 50 K and rapidly rise with further decrease in temperature. However, below $T \sim 30$~K, $H_{\rm c}$ starts to decrease which is contrary to that expected for a pure ferromagnetic phase. The only possibility for the observed modification of $H_{\rm c}$ could be the presence of other magnetic interaction around $T \leq T_{\rm L}$. So, it is most likely that the magnetic spin fluctuations associated with localized Nd-ions start to exert itself below 30 K and it is of competing character with the itinerant Co-3$d$ moment, resulting in the drop in $H_{\rm c}$. Also, $H_{\rm c}$ becomes almost temperature independent below $T_{\rm L}$ but a finite value remains even at the lowest measured temperature, which indicates a local antiferromagnetic ordering around $T_{\rm L}$ coexisting with finite ferromagnetic component down to 2 K. Thus, it is quite plausible that the Nd moments and Co moments settle for an short-range antiferromagnetic coupling below $T_{\rm L}$.

\subsection{\label{Sec:Cp} Heat capacity}

\begin{figure}[h!]
\centering
\includegraphics[width = 3.3in]{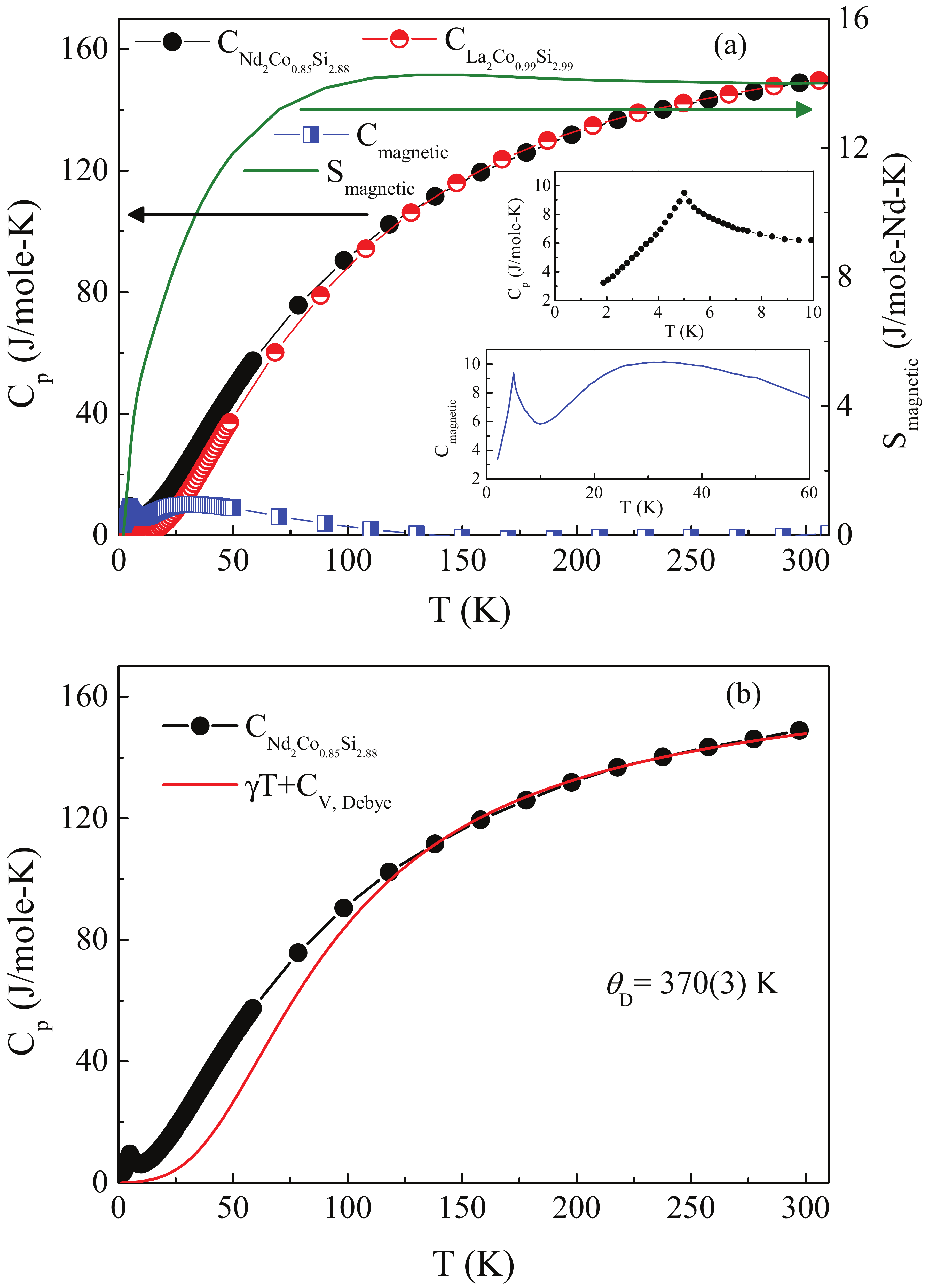}
\caption{(a) Zero-field heat capacity of Nd$_2$Co$_{0.85}$Si$_{2.88}$ and its non-magnetic analogue La$_2$Co$_{0.99}$Si$_{2.99}$. The magnetic contribution to the heat capacity (left panel) and magnetic entropy (right panel) as a function of temperature have been estimated by subtracting the non-magnetic contribution. (Top Inset) presents the experimental $C(T)$ data in the region of magnetic ordering. (Bottom Inset) presents the expanded view of magnetic contribution for a limited range of temperature. (b) Heat capacity of Nd$_2$Co$_{0.85}$Si$_{2.88}$ fitted using Eq. \ref{HC}. }\label{Nd_HC}
\end{figure}

To clarify the nature of ground-state magnetic interactions, zero-field heat-capacity measurements have been carried out for Nd$_2$Co$_{0.85}$Si$_{2.88}$ and its non-magnetic isostructural analogue La$_2$Co$_{0.99}$Si$_{2.99}$. The $C_{\rm p}(T)$ for both compounds are shown in Fig.~\ref{Nd_HC}(a). A peak around $T_{\rm L}$ is observed for Nd$_2$Co$_{0.85}$Si$_{2.88}$ on the onset of magnetic transition, although the height of the peak is much smaller than that observed in systems exhibiting a long-range magnetic ordering. The magnetic contribution to heat capacity ($C_{\rm {mag}}(T)$) for Nd$_2$Co$_{0.85}$Si$_{2.88}$ has been estimated by subtracting the lattice volume corrected \cite{anand2015antiferromagnetism} $C_{\rm p}(T)$ data of isostructural nonmagnetic analogue La$_2$Co$_{0.99}$Si$_{2.99}$. The $C_{\rm p}(T)$ data of Nd$_2$Co$_{0.85}$Si$_{2.88}$ and that of La$_2$Co$_{0.99}$Si$_{2.99}$ with lattice volume corrections perfectly match with each other for $T > T_{\rm H}$ in the phonon contribution dominated region. The estimated $C_{\rm {mag}}(T)$ clearly reflects the peak around $T_{\rm L}$ associated with the low-temperature magnetic transition in the system along with an additional broad hump-like anomaly having a maximum around 30 K (Fig.~\ref{Nd_HC}(a) inset). It is quite plausible that the high-temperature hump ($\sim 30$~K) might be associated with Schottky-type anomaly due to the likely presence of crystalline electric field (CEF) in the system. The estimated magnetic entropy at $T_{\rm L}$ is found to be about 60\% of the minimum value of magnetic entropy ($Rln2$) required for a system to be magnetically ordered having doublet spin configurations. Such a low value of $S_{\rm {mag}}$($T_{\rm L})$ suggest the absence of any true long-range magnetic ordering in the system and points towards a short-range ordering. In general, one can even argue that the estimation of magnetic entropy as 60\% of $Rln2$ might not be very accurate, as an inappropriate subtraction of nonmagnetic contribution or underestimation of low temperature ($T< 2$ K, which is beyond our measurement range) magnetic entropy could even make the magnetic entropy at $T_{\rm L}$ close to $Rln2$. It may however be prudent to compare the magnetic entropy of Nd$_2$Co$_{0.85}$Si$_{2.88}$ with that of isostructural Pr$_2$Co$_{0.86}$Si$_{2.88}$ where the magnetic entropy is estimated to be 48\% of $Rln2$ \cite{kundu2021complex}, very close to the estimated value of 60\% of $Rln2$ in Nd$_2$Co$_{0.85}$Si$_{2.88}$ . Since the Neutron diffraction study on Pr$_2$Co$_{0.86}$Si$_{2.88}$ does not indicate the presence of any long range magnetic order \cite{kundu2021complex}, a similar characteristic is also expected in Nd$_2$Co$_{0.85}$Si$_{2.88}$ as well. The non-equilibrium dynamical behavior (\textit{viz.}, magnetic-relaxation behavior, aging phenomena, and magnetic memory effect both in ZFC and FC protocols discussed in sec.~\ref{Sec:Neq}) further confirm the glassy-state formation below $T_{\rm L}$. Notably, quite a few members of this series with different transition elements are already reported to exhibit magnetically frustrated glassy behavior along with coexisting spatially limited magnetically ordered phase~\cite{pakhira2016large,pakhira2019spatially}. Such magnetically ordered states having small coherence length can also exist in this system, which remains a future study. Magnetic entropy saturates above 140 K at a value $\sim$~14 J/mole-Nd-K which is about 73\% of that expected for Nd$^{3+}$ ion [$Rln(2J+1)=Rln10$]. The lower estimation of magnetic entropy is in consonance with our earlier arguments of presence of short range type of magnetic ordering. As it is earlier reported that the magnetic correlation lengths in many other R$_2$NiSi$_3$-type of compounds are quite low (35-250 {\AA}) \cite{pakhira2019spatially,pakhira2016large,pakhira2017magnetic}, and some of them are also reported to consist of multiple magnetic structures with different $k$-vectors \cite{pakhira2016large, pakhira2021magnetic}, a considerable fraction of Nd ions in the intermediate regions of different magnetic grains are not susceptible to any magnetic order. Moreover, since Nd$_2$Co$_{0.85}$Si$_{2.88}$ form in non-stoichiometric defect structure, different Nd-ions would respond differently to the magnetic field depending on their local environment. As a result, some of the Nd-ions would remain paramagnetic down to the lowest temperature and would not contribute to the magnetic entropy of the system. A similar or comparable lower value of magnetic entropy has earlier been reported in a few other similar compounds, \textit{viz.}, Pr$_2$Co$_{0.86}$Si$_{2.88}$ ($0.76Rln9$) \cite{kundu2021complex}, Pr$_2$Ni$_{0.95}$Si$_{2.95}$ ($0.76Rln9$) \cite{pakhira2018unusual}, Tb$_2$Ni$_{0.90}$Si$_{2.94}$ ($0.87Rln13$) \cite{pakhira2019spatially}, etc.

The $C_{\rm p}(T)$ data has been analyzed using Debye model of heat capacity using the relation,
\begin{eqnarray}
C_{\rm p}(T)&=&\gamma T+C_{\rm D}(T) \label{HC}
\end{eqnarray}

\noindent where $\gamma T$ is the electronic contribution and the second term represents the Debye lattice heat capacity. $n$ is the number of atoms per formula unit. According to Debye model, $C_{\rm D}(T)$ could be written as

\begin{eqnarray}
% \nonumber % Remove numbering (before each equation)
C_{\rm D}(T)&=&9nR{\left(\frac{T}{\theta_{\rm D}}\right)}^3\int_{0}^{\theta_{\rm D}/T}\frac{x^4e^x}{(e^x-1)^2} dx\label{debye_HC}
\end{eqnarray}

\noindent where $x=\theta_{\rm D}/T$ and $\theta_{\rm D}$ is the Debye temperature. The fitting of heat capacity data using Debye model (Eq.~\ref{HC}) is shown in Fig.~\ref{Nd_HC}(b). The $C_{\rm p}(T)$ of Nd$_2$Co$_{0.85}$Si$_{2.88}$ could only be described using the Debye model for $T > T_{\rm H}$ with a Debye temperature of $\theta_{\rm D} = 370(3)$~K, which is close to that we have obtained from analysis of temperature dependent lattice volume parameters estimated from low temperature XRD patterns (Sec.~\ref{Sec:structure}). Similar values of $\theta_{\rm D}$ are also reported for different other members of R$_2$TX$_3$ series of compounds \cite{siouris2011antiferromagnetic,pakhira2018magnetic,iyer2007superconducting}.

\subsection{Computational Results}

\begin{figure}[h!]
\centering
\includegraphics[width = 3.5in]{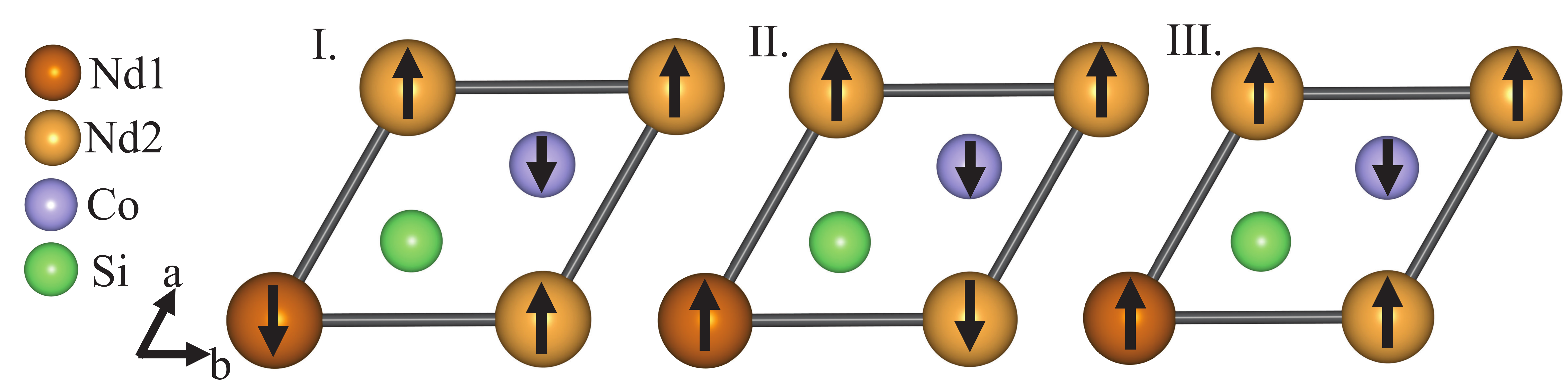}
\caption{Nd-4$f$ spin configuration in Nd$_2$CoSi$_3$ (pseudo-primitive): I. Nd1 and Nd2 are antiparallel, II. one Nd-atom at site-2 is antiparallel to others, III. Nd spins of both sites are parallel.}\label{plots-v2}
\end{figure}

\begin{figure*}
\centering
\includegraphics[width = 6in]{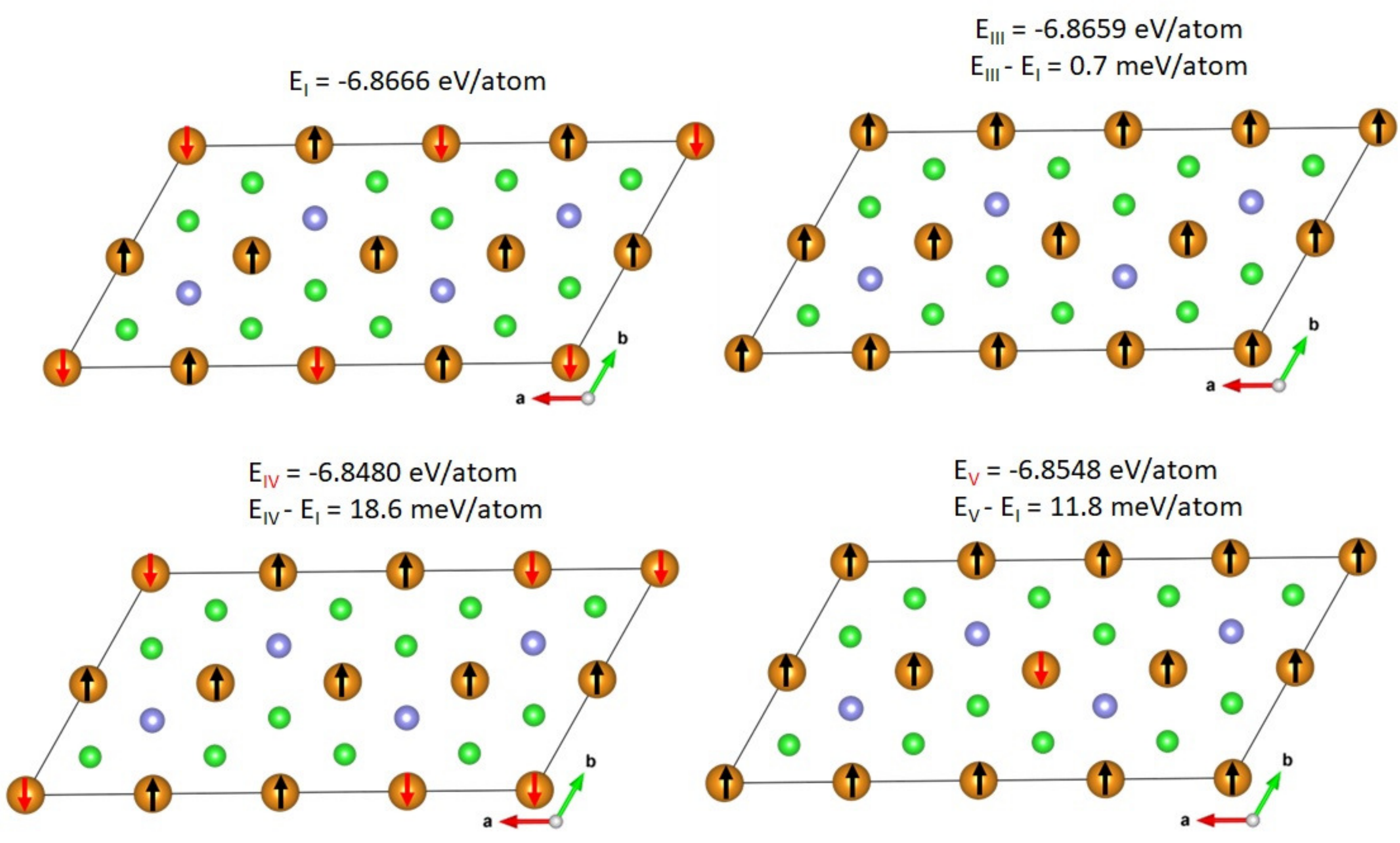}
\caption{Configurations I and III from the Fig.~\ref{plots-v2} are extended into 2$\times$1$\times$1-supercell. IV and V are newly added configurations.}\label{Fig_8}
\end{figure*}

As is well known, most rare-earth elements are in a 3+ state and, according to Hund's rule, their spin ($S$) and orbital ($L$) moments can be estimated easily. Nd in 3+ state will have 3 electrons in $f$-state such that $S$ = 2 $\times$ 3/2 = 3 $\mu_{\rm B}$ and $L$ = 6 $\mu_{\rm B}$, resulting in total moment of Nd$^{3+}$ ion = $6-3$ = 3 $\mu_{\rm B}$ without considering any SOC-effect (from Hund's rule: $L-S$ for light rare-earths and $L+S$ for heavy rare-earths). Due to the SOC effect, \textit{i.e.}, including g-factor, the total moment for Nd$^{3+}$ ion will be 3.62 $\mu_{\rm B}$. In materials, due to crystal-field effect/SOC, the orbital moment may get quenched. Our measured saturated magnetic moment of 4.17 $\mu_{\rm B}$/f.u. indicates a ferromagnetic coupling of two Nd atoms per f.u. -- probably with the quenched orbital moment, when Co-contribution is much weaker, as experimentally observed. Therefore, based on experimental results and basic knowledge of rare-earth ions, we have considered the configuration III along with two others (I and II: ferrimagnetic coupling of Nd atoms per cell) just by reorienting one spin to see the results of spin frustration and the possibility of ferrimagnetic coupling between Nd-atoms per cell. Nd$_2$CoSi$_3$ (the base f.u., which differs from cell that we use from Fig.~\ref{structure_paper}; Nd$_4$Co$_2$Si$_6$) has two Nd-sites; Nd1 is the second nearest-neighbor (2NN) of Co and Nd2 is the first NN of Co.

Here, three configurations (Fig.~\ref{plots-v2}) are considered as input: I. Nd1 and Nd2 are antiparallel; II. One of the Nd2-spins is antiparallel to the others; III. All Nd-atom spins are parallel. And, in all spin structures, Co-spin is antiparallel to average Nd-4$f$ spins. We find that spin configuration-I is lowest energetically (and stable - with a negative formation energy, see below). The energy difference ($\Delta E$) for a 12-atom unit cell (Fig.~\ref{structure_paper}) between configuration I ($E_0$ (I)$=-82.39939$ eV)  and configuration-II ($E_0$ (II)$=-82.39582$ eV per cell) and  III ($E_0$ (III)$=-82.39162$ eV per cell) are 1.79 meV/f.u.(or 0.30 meV/atom) and 3.89 meV/f.u. (or 0.65 meV/atom), respectively.

\begin{figure}
\centering
\includegraphics[width = 3.5in]{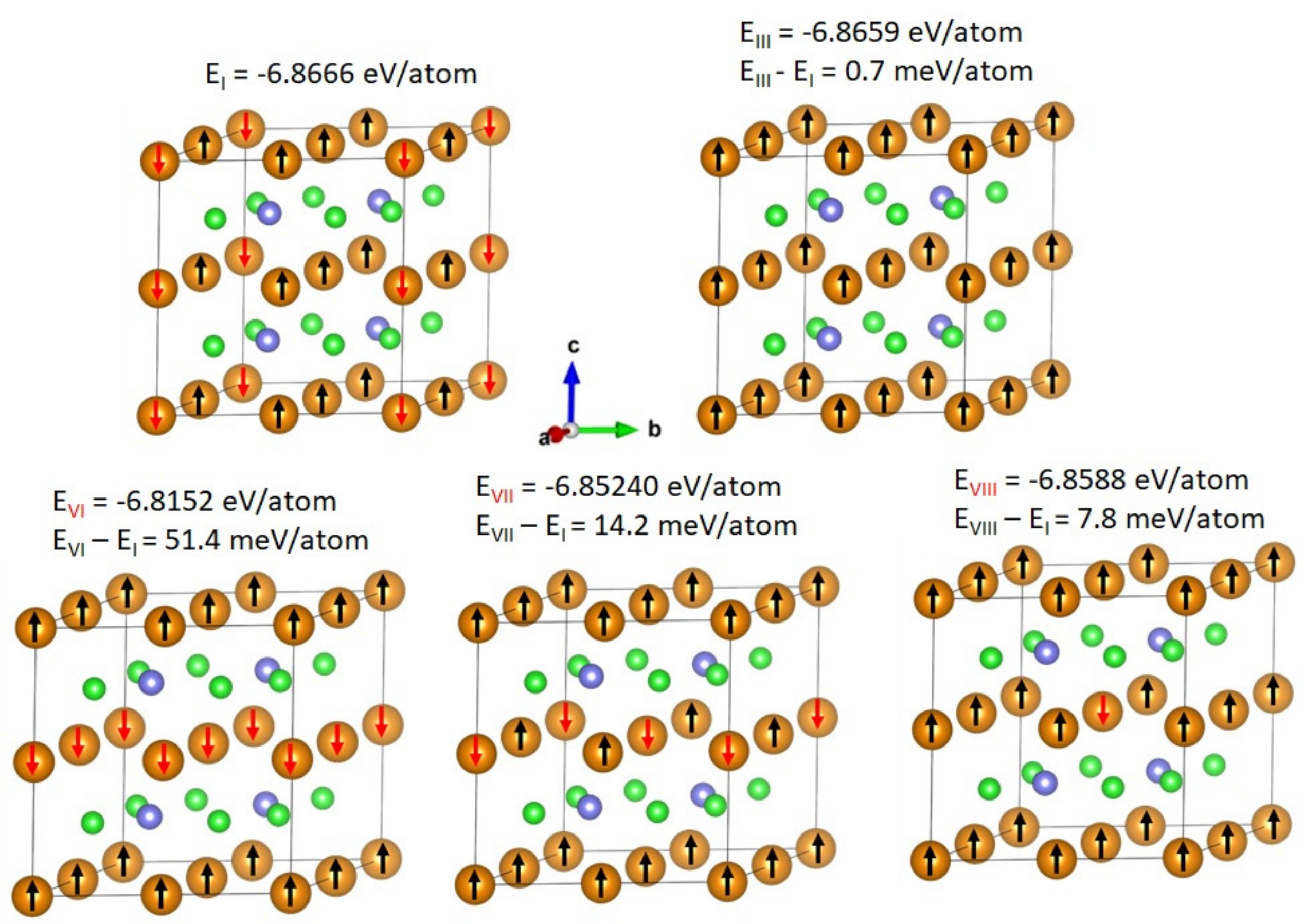}
\caption{Configurations I and III from Fig.~\ref{plots-v2} are extended into 1$\times$1$\times$2-supercell. VI, VII and VIII are newly added configurations}\label{Fig_9}
\end{figure}

To further support our interpretation, we have checked a few more configurations arranging Nd-spins ordered in an in-plane arrangement (IV-V; supercell 2$\times$1$\times$1) and an ordered arrangement along the $z$-axis (VI-VIII; supercell 1$\times$1$\times$2), see Figs.~\ref{Fig_8} and ~\ref{Fig_9}. For configurations with long-range order along the $z$-axis, having planes separated by Co,Si layer, the energies are 14-54~meV/atom higher, \textit{i.e.}, not competing. For a single moment flip, the energy is 8~meV/atom, still not competing with local disordered moments. Within a single Nd plane, configurations I and III (taken from Fig.~\ref{plots-v2}) are extended into the supercells and shown for comparison. After comparing the ground-state energies, we still have lower energies for configurations I-III. Even single planar antiparallel moment is 12~meV/atom higher. This supports the experiment that there is no long-range order amongst the Nd ions in this system. And furthermore, the Nd-spins are frustrated locally, as shown by the energetics.

\par For all three configurations (I, II, and III in Fig.~\ref{plots-v2}), we determined the formation energy, $\Delta E_{form}$, of Nd$_2$CoSi$_3$ (bulk without vacancy), defined by:
\begin{equation}
\Delta E_{form}=E_0-4\times \mu_{\rm Nd}-2\times \mu_{\rm Co}-6\times \mu_{\rm Si},
\label{formation_energy}
\end{equation}
where $E_0$ is total ground-state energy of a bulk spin-configuration per cell of Nd$_4$Co$_2$Si$_6$, and $\mu_{\rm Nd}$, $\mu_{\rm Co}$ and $\mu_{\rm Si}$ are the chemical potentials from pure bulk Nd, Co and Si, respectively. For thermodynamic purposes, the $\Delta E_{form}$ is reported per atom, where a negative value indicates the stability of the systems relative to the pure end point elemental forms. We find the formation energy in Eq.~\ref{formation_energy} for configuration I is lowest at $–0.78133$ eV/atom, but configuration II ($–0.78103$ eV/atom) and III ($–0.78068$ eV/atom) are higher in energy by less than 1 meV/atom, specifically, 0.30 and 0.65 meV/atom, respectively, the same as from cell energy differences. Energies differ for configurations I and II (III) by less than 7 K.

\par Magnetic frustration is common in spin glasses and linked to multiple ground states. Theory results support the magnetic frustration as the energy difference between the considered configurations is very small. As such, there FM and AFM spin coupling between nearest-neighbours are competing, as expected in a spin glass. With all three configurations competing [II (III) is 1.8 (3.9) meV/f.u. higher than I, as found above], we find that configuration III has a total moment close to the measured saturated moment-value, suggesting that the atomic arrangements are in a configuration-III state with the canted spin and quenched orbital moment. The atomic and total spin-moment values (with and without SOC) in each configuration are given in Table~\ref{moment_theory}. The orbital magnetic moment ($L$) appears as a manifestation of SOC within DFT \cite{steiner2016calculation,solovyev1998hund,knopfle1997symmetry}. As such, the orbital moment reported in the Table contains the effect of SOC.  Hence, in DFT, the total moment is the addition of spin and orbital moment that directly correlates with the saturated moment in experiment.

\begin{table*}
\caption{Spin-polarized and SOC-moments (atomic \& total) in each spin-structure of Nd$_2$CoSi$_3$ in Fig.~\ref{plots-v2}. The arrows ($\uparrow$, $\downarrow$) represent the direction of respective in-plane $L/S$-spins.}

\begin{center}
\renewcommand{\arraystretch}{1.25}
%\begin{ruledtabular}

\begin{tabular}{|c | c c c | c c |}
\hline\hline

 & \multicolumn{3}{|c|}{Spin-polarized moment} & \multicolumn{2}{|c|}{SOC $L/S$-moment}\\
\cline{2-6}
Configuration & Nd-$S$ & Co-$S$ & Total $S$ & Nd1/Nd2-$L$ & Total-$S/L$\\
\cline{2-6}
& ($\mu_{\rm B}$/atom) & ($\mu_{\rm B}$/atom) & ($\mu_{\rm B}$/f.u.) & ($\mu_{\rm B}$/atom) & ($\mu_{\rm B}$/f.u.)\\
\hline\hline
I & 3.03 & 0.24$\downarrow$ & 2.79 & 4.1$\uparrow$/2.7$\downarrow$ & 2.75/2.04\\
II & 3.02 & 0.09$\downarrow$ & 2.90 & 2.93$\downarrow$/3.34$\uparrow$~2.8$\downarrow$ & 2.87/2.51\\
III & 3.03 & 0.24$\downarrow$ & 5.78 & 3.51$\downarrow$/5.38$\downarrow$ & 5.77/9.83\\
\hline\hline
\end{tabular}
\end{center}
\label{moment_theory}
\end{table*}

\par The Nd $S$-moments with and without SOC are the same as expected. The $L$-moment of structure-I for Nd1 and Nd2 are 4.1 and 2.7 $\mu_{\rm B}$ per f.u., respectively. The total moment $\lvert$L-S$\rvert$ of structure-I is 0.71 $\mu_{\rm B}$ per f.u. In this case, $L$-moment of 2.04 $\mu_{\rm B}$ per f.u. is quenched significantly from the Hund's value 6 $\mu_{\rm B}$ per f.u. In structure-II, where Nd2-4$f$ spins are ferrimagnetic, the $L$-moment of each atom on site-2 is different. $L$-moments with ``down" direction are quenched. Whereas $S$- and $L$-moments of structure-III are 5.78 and 9.83 $\mu_{\rm B}$ per f.u., respectively, result in a total moment of 4.05 $\mu_{\rm B}$ per f.u. Quenching of moments in this structure is smaller; Nd1 and Nd2 have orbital moments of 3.51 and 5.38 $\mu_{\rm B}$ per atom, respectively. The Co-moment is antiparallel in all structures (similar to the case without SOC). Also, the $S$- and $L$-moment of Nd-atoms are antiparallel with each other.

\par Figure~\ref{plots-v2} shows the initial configurations considered for calculations, which converged to canted (random orientation) Nd-spins in xy-plane; more prominent for structure I and II and less so in III. The moment values, see Table~\ref{moment_theory}, are the resultant of moments in x and y-direction. Therefore, we predict that all spin-structures are competing at low temperatures, indicating spin frustration; an anomaly in the experimental magnetic susceptibility (Fig.~\ref{Nd_MT}) and heat capacity (Fig.~\ref{Nd_HC}) at low temperature is due to a reduction in the moment, especially, quenching the orbital moment along with canted spins. In rare-earth (R)–transition-metal (T)-based systems like SmCo$_5$, quenching of orbital moment happens when electrostatic crystal field (CF) energy competes with SOC energy \cite{das2019anisotropy}. The SOC affects the R-4$f$ prolate charge, which is the origin of $L$-quenching.

\subsubsection{Vacancy Defects}
We also investigated the effect of vacancies on each configuration by adding 12.5\% vacancies on Co sublattice sites and 4.167\% vacancies on the Si sublattice sites, yielding a supercell composition (Nd$_2$Co$_{0.875}$Si$_{2.88}$) close to that experimentally observed (Nd$_2$Co$_{0.85}$Si$_{2.88}$). We constructed $2\times 2\times 1$-supercell of the Nd$_4$Co$_2$Si$_6$ cell and removed two atoms (a Si and Co) for defect-modified composition and considered spin-configuration I and III for comparison. The ground state-energy differences of these configurations with divacancies is 2.4 meV/atom, compared to 0.65 meV/atom without vacancies given above.

\par A formation energy for a defected cell is calculated in a supercell as \cite{zhang2018calculating}:
\begin{equation}
E_f^{defect}=[E_{defect}-E_o (N)]\pm \sum\Delta n_i\mu_i
\label{energy_defect}
\end{equation}

\noindent where $E_{defect}$ and $E_o (N)$ are the total energies of system with and without defects, respectively, $\Delta n_i$ represents the number of atoms of type `$i$' which are added ($\Delta n_i<0$; interstitials) or removed ($\Delta n_i>0$; vacancy) to make a defective system, and $\mu_i$ is the corresponding chemical potential associated with each defect atom. Here, we remove one Co and one Si atom from a $2\times 2\times 1$ supercell composed of four Nd$_4$Co$_2$Si$_6$ cell (as in Fig.~\ref{structure_paper}) to make a divacancy supercell having 48 atom sites total, \textit{i.e.}, Nd$_{16}$Co$_7$Si$_{23}$. The divacancy formation energy is then:
\begin{equation}
E_f^{vac}= [E_{vac}-E_0 ]+ \mu_{Co}+ \mu_{Si}
\label{energy_vacancy}
\end{equation}
where $E_0$ is the bulk (0 vacancy) energy (4$E_{\rm Nd_4Co_2Si_6}$ for the supercell) and $E_{vac}$ is the divacancy energy for the Nd$_{16}$Co$_7$Si$_{23}$ supercell. Again, $\mu_{\rm Co}$ and $\mu_{\rm Si}$ are the chemical potentials of Co and Si elements as their ground-state energies, taken from the Materials Project database~\cite{jain2011high, jain2013commentary}. Notably, the chemical potential of elements is equal to the DFT total energy of their ground states in most cases, but, in some cases, due to over/under-binding nature of elements in the solids, that gives an error in formation energies, \textit{e.g.}, O$_2$ molecule in oxides. Therefore, chemical potentials reported in the database are corrected based on comparison of experiment and theory, according to the scheme in refs.~\cite{wang2021framework, wang2006oxidation, jain2011formation}.

\par The $E_f^{vac}$ for configuration I (III) is 1.75 (1.79) eV per vacancy. If we compare the ground-state energies of parent ($E_0$) and defective supercells ($E_{vac}$), the difference ($E_{vac}-E_0$) for spin-configuration I (III) is found to be 0.050 eV/atom (0.052 eV/atom). This suggests local magnetic states are affected by the presence of vacancies, but the overall electronic structure is little affected (that is, $E_f^{vac}$ is roughly the same for I and III).

\subsubsection{Electronic Properties}

\begin{figure}[h!]
\centering
\includegraphics[width = 3in]{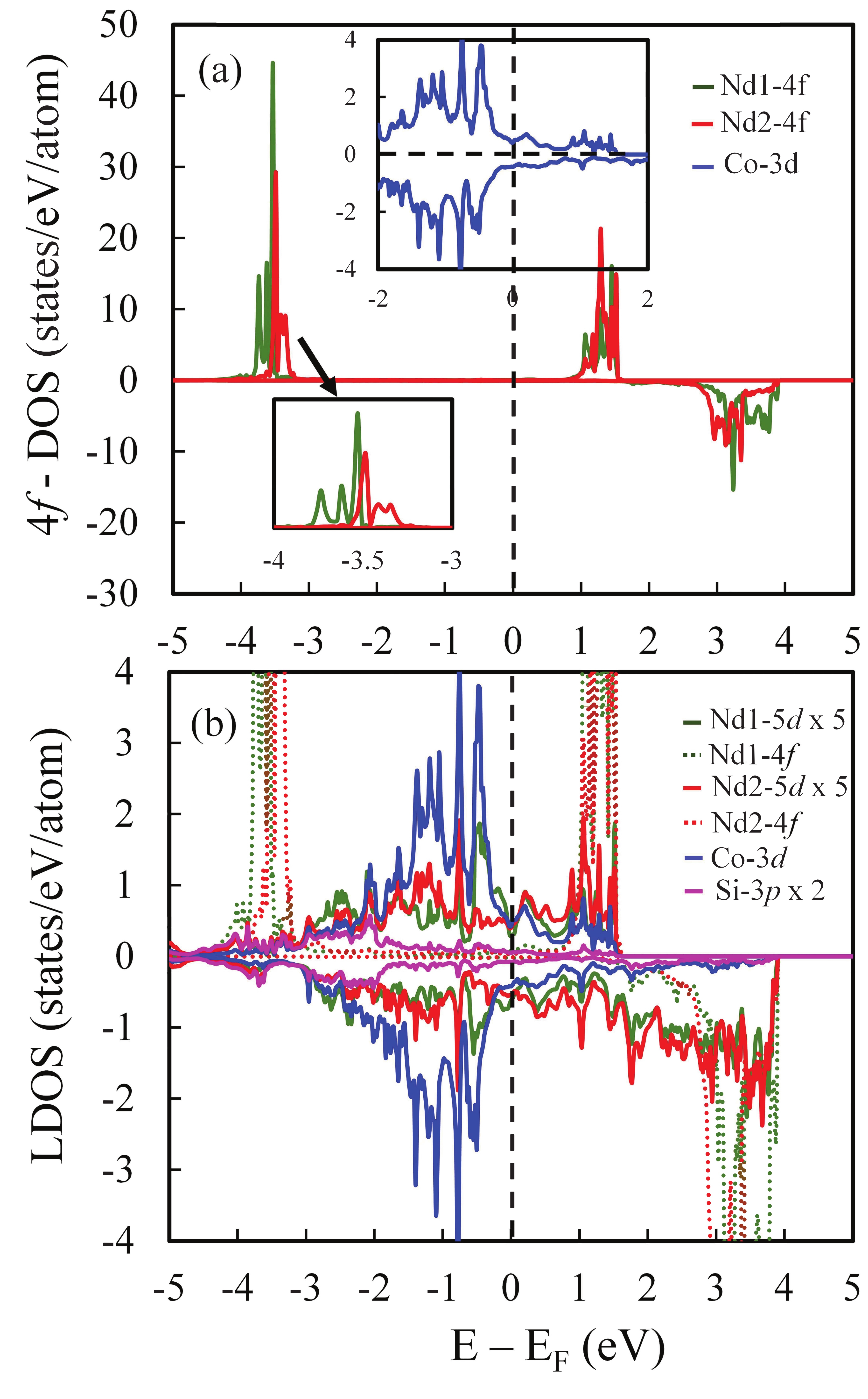}
\caption{Spin-polarized local density of states (LDOS) of Nd$_2$CoSi$_3$ with spin structure-III (see Fig.~\ref{plots-v2}) for (a) 4$f$ and 3$d$ states of Nd and Co, (b) 3$p$, 3$d$, 5$d$, and 4$f$ states showing hybridization. Note: Nd-5$d$ and Si-3$p$ states are magnified by a factor of 5 and 2, respectively, in the figure.}\label{fig_DOS_theory}
\end{figure}

To explore the electronic structure, the spin-polarized local density of states (LDOS) of spin structure-III (whose magnetic moment matches with the observed value) is shown in Fig.~\ref{fig_DOS_theory}. The Nd-4$f$ DOS at both sites and Co-3$d$ DOS in the inset (top) are shown in Fig.~\ref{fig_DOS_theory}(a). 4$f$-states are well localized and split into three peaks in valence-band around $-3.5$ eV (zoom part is in the inset), representing magnetic quantum number, $m = +3, +2, +1$ \textit{i.e.} Nd$^{3+}$ ion. Conduction-band peaks are collection of $m = 0, -1, -2, -3$. On the other side, Fig.~\ref{fig_DOS_theory}(b) shows the atomic hybridization in the system. Si-3$p$ and Nd-5$d$ states are magnified by a factor of 2 and 5, respectively to highlight the hybridization. A strong hybridization is predicted between Co-3$d$ and Nd-5$d$ states in the valence band and at the Fermi energy (E$_{\rm F}$), there is a valley in the majority channel, making a pseudo-gap. In addition, Co-3$d$ states have no spin-polarization at the E$_{\rm F}$ which is highly spin-polarized in FM Co-atom \cite{ziese2007spin}, and the majority and minority spin states have almost equal weightage in the system. The cobalt behavior in the system is like a Cu-atom as its DOS is very close to Cu's \cite{ziese2007spin}. Diminished spin-polarization of Co causes the small Co-moment. Interestingly, Nd-5$d$ states, hybridizing with Co-3$d$ states, also have missing spin polarization at E$_{\rm F}$. Also, the weightage of states from both channels is almost equal results in almost zero (5$d$-moment $\sim$ 0.01 $\mu_{\rm B}$ per atom) contribution from Nd-5$d$ states. This indicates that 3$d$/5$d$-conduction carriers are blocked in this system and act as a barrier for the 4$f$-4$f$ interactions, resulting in spin-frustration.

\par The magnetic moment of the metal is correlated directly with the exchange-splitting of the DOS at the Fermi energy~\cite{kozlenko2015sequential}. Therefore, the lack of exchange splitting in the DOS supports the small Co-moment (0.24 $\mu_{\rm B}$). In the Stoner model, collapsing of exchange splitting (or quenching of magnetic moment of itinerant metal) may be because of single-particle excitations whereas, in the Heisenberg model, local spins can exhibit collective excitations, \textit{i.e.}, spin fluctuation or spin canting~\cite{eich2017band}. No canting is found for Co moments, it is quenched due to a collapse of exchange splitting, indicating the itinerant nature of Co-moment.

\par Chemical hybridization, especially in the vicinity of the Fermi energy, is an important factor. The electron localization is the tendency of electrons to be confined in a small region limiting localized states from being involved in chemical hybridization with the other states. In the present system, Co-3$d$ states are strongly hybridized with Nd-5$d$ states near to the Fermi energy, additional supporting the itinerant nature of the Co moments.

\subsection{\label{Sec:Neq} Non-equilibrium dynamics}

Quite a few members of R$_2$TX$_3$ type indeed exhibit glassy feature associated with localized rare-earth spins \cite{pakhira2016large,tien1997mass,mo2015magnetic,majumdar2001multiple}. The magnetization, heat capacity data and DFT calculations reveal the presence of strong spin frustration in the system as a competition of localized Nd and itinerant Co moments. In the presence of such frustration and inherent bond-disorder in the system, the low temperature magnetic state is expected to undergo spin-glass type of behavior below $T_{\rm L}$. The possible presence of such glassy magnetic phase in Nd$_2$Co$_{0.85}$Si$_{2.88}$ is examined through a comprehensive study on the low-temperature non-equilibrium dynamical behavior.

\begin{figure}[h!]
\centering
\includegraphics[width = 3.2in]{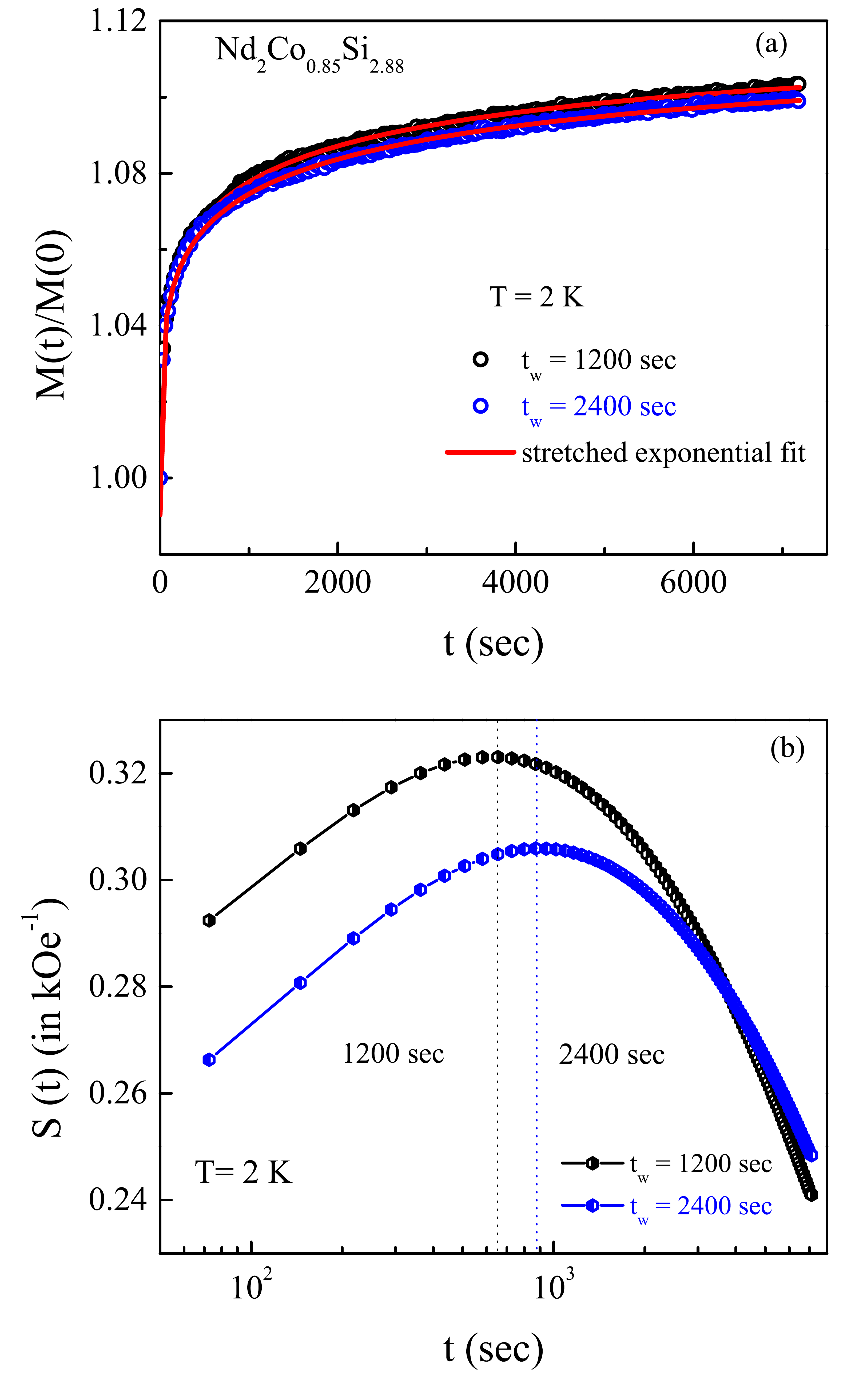}
\caption{(a) Time dependent normalized magnetization of Nd$_2$Co$_{0.85}$Si$_{2.88}$ at $T$ = 2 K under ZFC protocol for different wait times, along with stretched exponential fit (Eq.~\ref{stretchedexponential}). (b) The relaxation rate $S(t)$ at $T$ = 2 K for two different wait times in ZFC protocol for Nd$_2$Co$_{0.85}$Si$_{2.88}$, exhibiting aging phenomenon.}\label{Nd_relax_diff_wait_time}
\end{figure}

Magnetic relaxation behavior is a key characteristic feature of glassy state formation in a system. Such relaxation behavior can generally be measured in both ZFC and FC protocols in the dc magnetic measurements. Here, the relaxation measurements have been carried out in ZFC protocol to avoid any possible negative impact of residual magnetic field in the commercial superconducting magnetometer. In ZFC procedure, the sample was cooled down from paramagnetic region to the measurement temperature (2 K $< T_{\rm L}$) in the absence of any magnetic field. After the temperature is stabilized for a wait time $t_{w}$, a small amount of magnetic field ($\mu_0{H}< k_B T_{\rm L}$) was applied followed by recording $M(t)$ for a long time scale, as shown in Fig.~\ref{Nd_relax_diff_wait_time}(a) for two different $t_{w}$. The presence of magnetic relaxation behavior clearly indicates the presence of magnetic glassy component in the system below $T_{\rm L}$. The time-dependent magnetic relaxation for a magnetically frustrated system is well described by \cite{mydosh1993spin,binder1986spin},

\begin{equation}
\ M({t}) = M_{0} \pm M_{g}{\rm exp}{\left[-\left({\frac{{\rm t}}{\tau}}\right)^{\beta}\right]}
\label{stretchedexponential}
\end{equation}

\begin{figure}[h!]
\centering
\includegraphics[width = 3.2in]{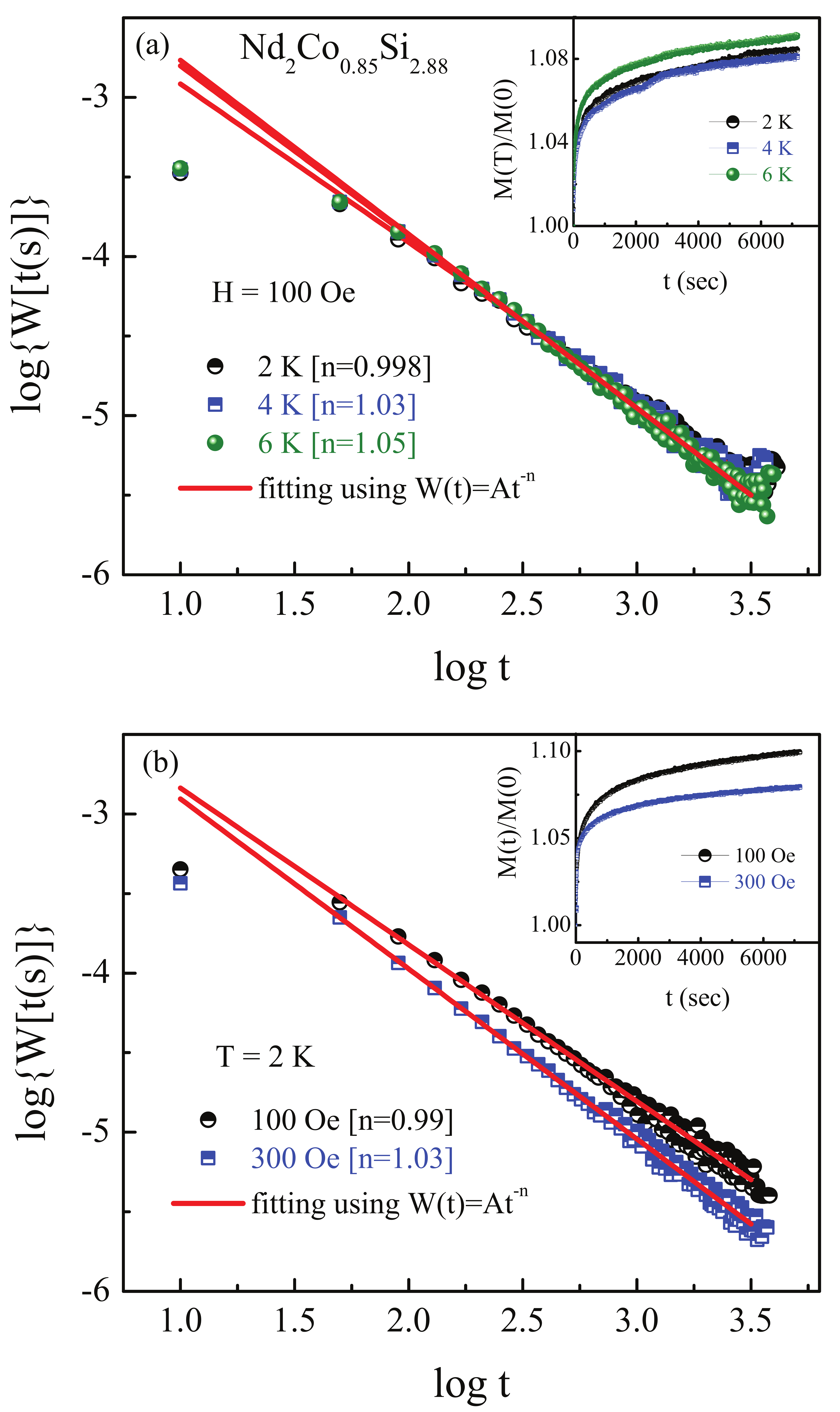}
\caption{(a) Magnetic relaxation rate as a function of time on a log-log plot for Nd$_2$Co$_{0.85}$Si$_{2.88}$ at different temperatures obtained from the corresponding ZFC magnetic relaxations in a 100 Oe field as shown in the inset. The straight lines correspond to the linear fit following Eq.~\ref{Ulrich} and the exponent values n are determined from the slopes of the fit. (b) Time dependence of magnetic relaxation rate on a log-log plot at 2 K for different magnetic fields, obtained from corresponding ZFC magnetic relaxations as shown in the inset. The exponent values n for different fields are determined from the linear fits (Eq.~\ref{Ulrich}) to the relaxation rate.}\label{Nd_relax_diff_TH}
\end{figure}

\noindent where $M_0$ denotes intrinsic magnetization present in the compound, $M_g$ is the glassy component of magnetization, $\tau$ is the relaxation time constant and $\beta$ is known as the stretching exponent. The value of $\beta$ depends on the nature of energy barriers involves in the relaxation process. $\beta$ = 0 implies no relaxation and $\beta$ = 1 is for single time constant relaxation process. The value of $\beta$, which ranges between 0 and 1, indicates the glassy nature of a system as well as it gives an estimation of the distribution of energy barriers present in the frozen state. The fitted $M(t)$ data of Nd$_2$Co$_{0.85}$Si$_{2.88}$ for different wait times $t_{w}$ using Eq.~\ref{stretchedexponential} is shown in Fig.~\ref{Nd_relax_diff_wait_time}(a). It is found that $\beta$ value is 0.29(1) for both the cases which is in the same range observed for different other glassy systems reported earlier~\cite{mydosh1993spin,chu1994dynamic} and also in the same range observed in R$_2$NiSi$_3$ systems \cite{pakhira2016large,pakhira2018unusual,pakhira2018chemical,pakhira2018magnetic}. The estimated values of $\tau$ is found to increase with increase in $t_{w}$ reflecting the fact that the system memorized the information about the wait time before the relaxation process starts. This phenomena is known as aging process for glassy systems. The aging phenomena can also be exemplified by fitting the relaxation rate

\begin{equation}
\ S(t) = \frac{1}{H}\frac{dM(t)}{d({\rm log} t)},
\label{viscosity}
\end{equation}

\noindent which could be obtained from the logarithmic time derivative of the ZFC susceptibility \cite{campbell2013relaxation}. The time dependent behavior of $S(t)$ for Nd$_2$Co$_{0.85}$Si$_{2.88}$ is shown in Fig.~\ref{Nd_relax_diff_wait_time}(b). The inflection point in $M(t)/M(0)$ curves corresponds to a maximum in $S(t)$. This maximum shifts to longer observation time with increasing $t_{w}$ implying continuous aging of the system below $T_{\rm L}$. The temperature and magnetic field evolution of the magnetic relaxation has been further investigated by measuring $M(t)$ as a function of time (under ZFC condition), at different temperatures and applied magnetic fields (Fig.~\ref{Nd_relax_diff_TH}). According to a theoretical model proposed by Ulrich \textit{et al.} \cite{ulrich2003slow}, the rate of change of normalized remanent magnetization $W(t) = - (d/dt)[{\rm ln}m(t)]~ ({\rm where} m(t)$= $M(t)$/$M(0)$) for a system consisting of interacting magnetic clusters decays following the law,

\begin{equation}
\ W(t) = At^{-\rm n}, t\geq t_0,
\label{Ulrich}
\end{equation}

\noindent where $A$ is a constant, n is an exponent function of temperature, and $t_0$ is the crossover time. The value of n indicates the dipolar interaction strength among the magnetic clusters present in the system. For dilute systems, dipolar interaction is negligible but it becomes more important for systems with increasing density. The relaxation rates $W$ of Nd$_2$Co$_{0.85}$Si$_{2.88}$ as a function of time $t$ in a log$_{10}$-log$_{10}$ plot for different temperatures and different magnetic fields are shown in Figs.~\ref{Nd_relax_diff_TH}(a) and (b), respectively. The value of n estimated from the fit of $W$ using Eq.~\ref{Ulrich}, is found to be close to 1, indicating strong dipolar interaction. The value of n is sensitive to both temperature and applied magnetic field, signifies spin-cluster glass behavior \cite{pakhira2016large,manna2013correspondence,rivadulla2004origin} of the present system at low temperatures.

\begin{figure}[h!]
\centering
\includegraphics[width = 3.2in]{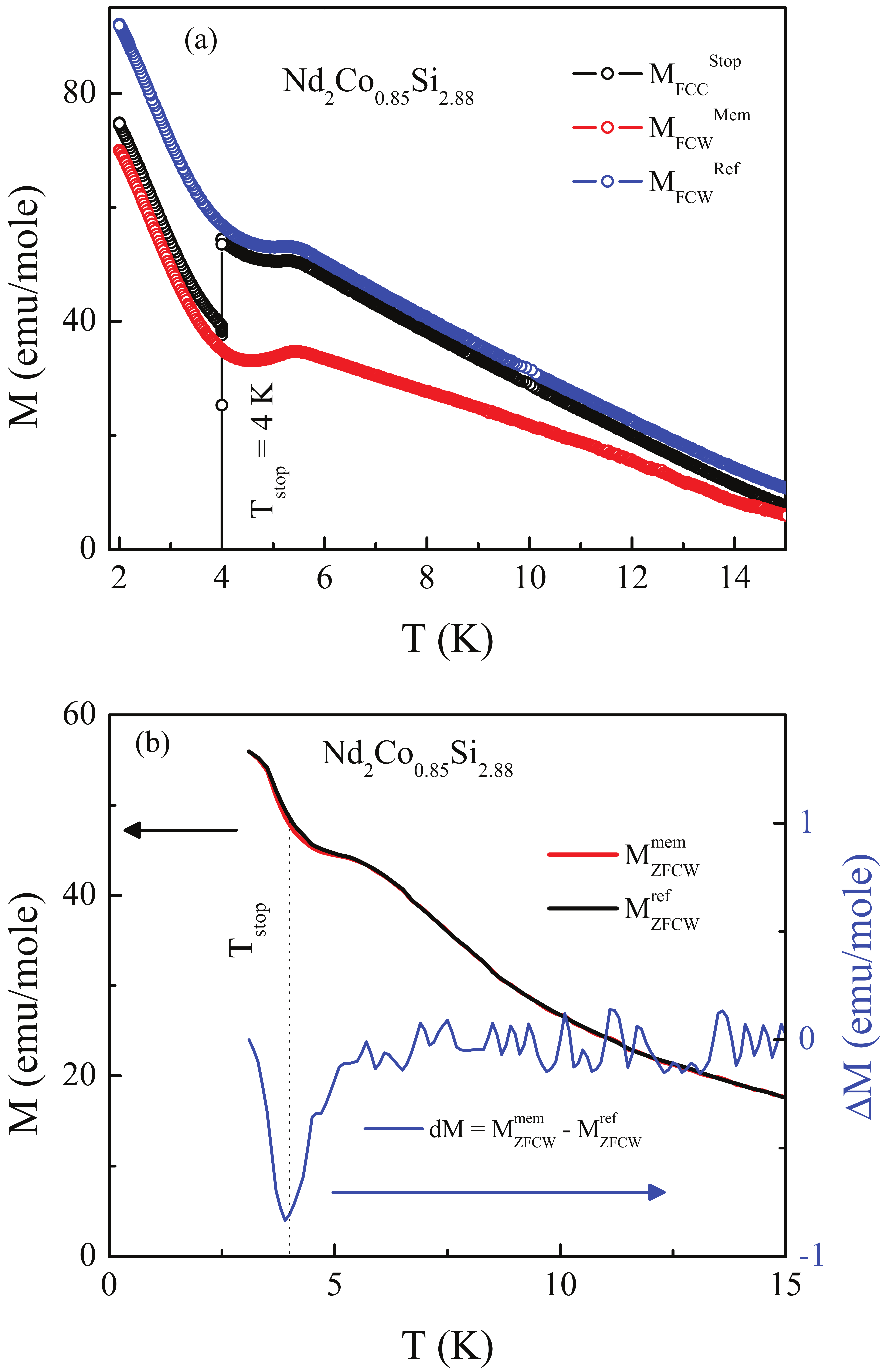}
\caption{Magnetic memory effect of Nd$_2$Co$_{0.85}$Si$_{2.88}$ in (a) field-cooled (FC) condition and (b) zero-field-cooled (ZFC) condition for 100 Oe applied field.}\label{Nd_memory}
\end{figure}

\begin{figure*}[ht!]
\centering
\includegraphics[width = 5.6in]{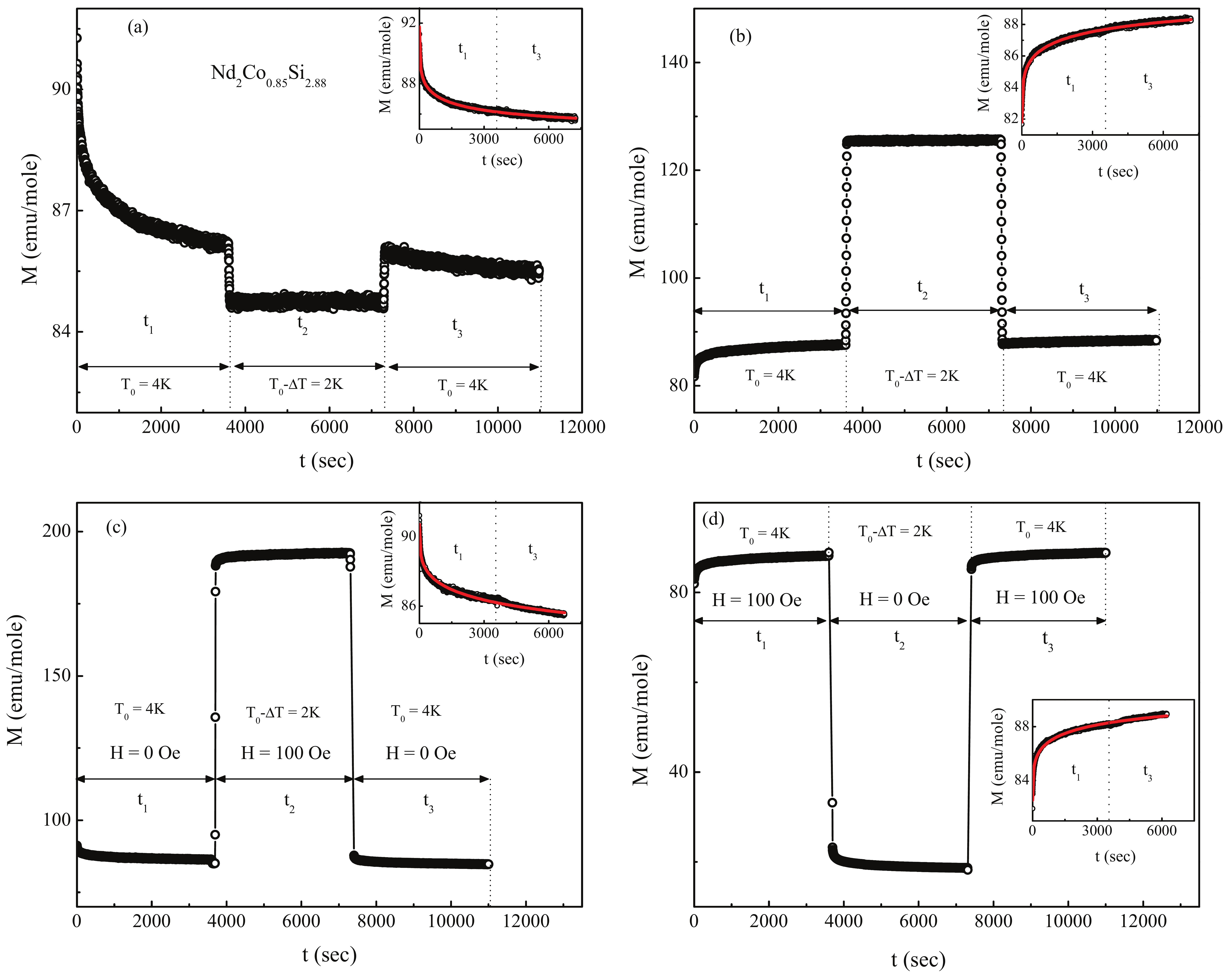}
\caption{Magnetic relaxation behavior of Nd$_2$Co$_{0.85}$Si$_{2.88}$ at 4 K for $H$ = 100 Oe with temporary cooling at 2 K in (a) FC method and (b) ZFC method. Magnetic relaxation behavior at 4 K with an opposite relaxation during temporary cooling at 2 K using (c) the FC and (d) ZFC protocol. The insets present the relaxation data as a function of total time spent at 4 K along with the fit (solid red line) using stretched exponential function [Eq.~\ref{stretchedexponential}].}\label{Nd_tempcooling}
\end{figure*}

Magnetically frustrated glassy systems exhibit another most spectacular manifestation of the non-equilibrium dynamics below blocking temperature, known as magnetic memory effect \cite{sun2003memory}. Memory effect has been investigated for Nd$_2$Co$_{0.85}$Si$_{2.88}$ in both FC and ZFC conditions, following the protocol introduced by Sun {\it {et al}} \cite{sun2003memory}. In FC protocol, the sample was cooled from room temperature 300 K to the base temperature 2 K in the presence of 100 Oe magnetic field with a intermittent stop at $T_{\rm {stop}} = 4$~K ($< T_{\rm L}$), where the magnetic field was switched off allowing the system to relax for $t_{ w} = 1$~h amount of time. After the time lapse of $t_{ w}$, same amount of magnetic field was switched on again followed by resumed cooling. Magnetization measured in this process is depicted as $M_{\rm FCC}^{\rm stop}$. After reaching to 2 K, the sample was warmed up to room temperature with the same magnetic field along with the same rate of temperature sweep and the measured magnetization is presented as $M_{\rm FCW}^{\rm mem}$. A reference curve ($M_{\rm FCW}^{\rm ref}$) was also measured for the compound which is the conventionally measured FC magnetization curve. The magnetic memory effect measured in the FC protocol, for the compound is shown in Fig.~\ref{Nd_memory}(a). The magnetization curve shows that $M_{\rm FCW}^{\rm mem}$ tries to follow the step-like $M_{\rm FCC}^{\rm stop}$ curve at the $T_{\rm stop}$, exhibiting memory effect, similar to different spin-glass systems reported earlier to exhibit this kind of memory effect \cite{khan2014memory,bhattacharyya2011spin,jonason1998memory}. As a phase-separated or superparamagnetic systems are also known to exhibit memory effect in the FC process \cite{sasaki2005aging}, it is indeed necessary to investigate the magnetic memory effect in ZFC protocol. The true magnetically frustrated glassy systems are known to exhibit memory effect in ZFC condition too, however it is forbidden in other types of non-equilibrium systems. In the ZFC protocol, the sample was zero-field-cooled from the paramagnetic region to the stopping temperatures ($T_{\rm stop}$) in the absence of magnetic field, where the cooling was temporarily stopped for $t_{w}$ = 1.5 h. The cooling was then resumed down to the lowest temperature 2 K. After reaching the lowest temperature, a small amount of magnetic field (100 Oe) was applied and the magnetization was recorded during heating, designated as $M_{\rm ZFCW}^{\rm mem}$. The conventional ZFC magnetization $M_{\rm ZFCW}^{\rm ref}$ was also measured at $H = 100$ Oe. The obtained data measured in this protocol for the studied compound is shown in Fig.~\ref{Nd_memory}(b). The difference between the reference curve and the curve with aging, $\Delta M (= M_{\rm ZFCW}^{\rm mem} - M_{\rm ZFCW}^{\rm ref})$ exhibit memory dip at $T_{\rm stop}$. This phenomena indicates that the system remembers the temperature where the cooling process was temporarily stopped and gives another confirmation of true glassy state formation. Usually, one can explain memory effect in different glassy systems within the framework of two different theoretical models, \textit{viz.}, the droplet model \cite{fisher1988nonequilibrium,fisher1988equilibrium,mcmillan1984scaling} or the hierarchical model \cite{dotsenko1985fractal,vincent1997slow}. According to droplet model, both the heating and cooling cycles of magnetic relaxation are symmetric, as a system favours only one equilibrium spin configuration at a given temperature. On the other hand, the hierarchical model predicts a multivalley structure \textit{i.e.} an infinite number of metastable states separated by barriers in the free-energy landscape. During temporary cooling, each metastable state splits into new substates while these states merge into new states with temporary heating. Thus, according to hierarchical model, a system supports asymmetric behavior in magnetic relaxation upon cooling and heating. We have investigated the influence of temperature and field cycling on the behavior of magnetic relaxation in both the ZFC and FC methods following the protocol of Sun {\it {et al}} \cite{sun2003memory}. In ZFC mode, the sample was cooled with a constant cooling rate in the absence of magnetic field, from paramagnetic region to a temperature $T_0$ which is below $T_{\rm L}$. After reaching $T_0$, a small magnetic field is applied and magnetization $M(t)$ was recorded for $t_1$ = 1 h. Then the sample was cooled to a lower value $T_0$ - $\Delta T$ in the same field and $M(t)$ was measured for a time $t_2$ = 1 h. Finally, the sample temperature was restored to $T_0$ and $M(t)$ was measured for $t_3$ = 1 h. In FC process, the sample was initially cooled to $T_0$ from paramagnetic region in the presence of a small magnetic field. Once $T_0$ was reached, the magnetic field was switched off and subsequently $M(t)$ was measured following temperature cycle similar to ZFC. The FC and ZFC relaxation process with temporary cooling for the present system is shown in Fig.~\ref{Nd_tempcooling}(a) and (b), respectively. From the measurements it is found that in both ZFC and FC processes, the state of the system is preserved even after a temporary cooling. One may notice that the relaxation curve during $t_3$ seems to be a smooth continuation of that during $t_1$, and as a whole it can be fitted with a single curve following stretched exponential form of Eq.~\ref{stretchedexponential}. Such a restoration of the original spin configuration suggest the presence of memory effect in these compounds. We have also investigated the opposite relaxation behavior of the samples by switching on and off the applied field alternatively, to check the strengthness of memory effect. In FC method, field was off during $t_1$ and $t_3$ and while a magnetic field is applied during $t_2$ and $M(t)$ was measured throughout, as shown in Fig.~\ref{Nd_tempcooling}(c). In ZFC process, $M(t)$ was measured in the presence of magnetic field during $t_1$ and $t_3$, while field remain switched off during $t_2$. The related curve is shown in Fig.~\ref{Nd_tempcooling}(d) that shows that instead of such opposite relaxation during $t_2$, magnetic relaxation during $t_1$ and $t_3$ are almost continuous and can be well described using Eq.~\ref{stretchedexponential}.

\begin{figure}[h!]
\centering
\includegraphics[width = 3.2in]{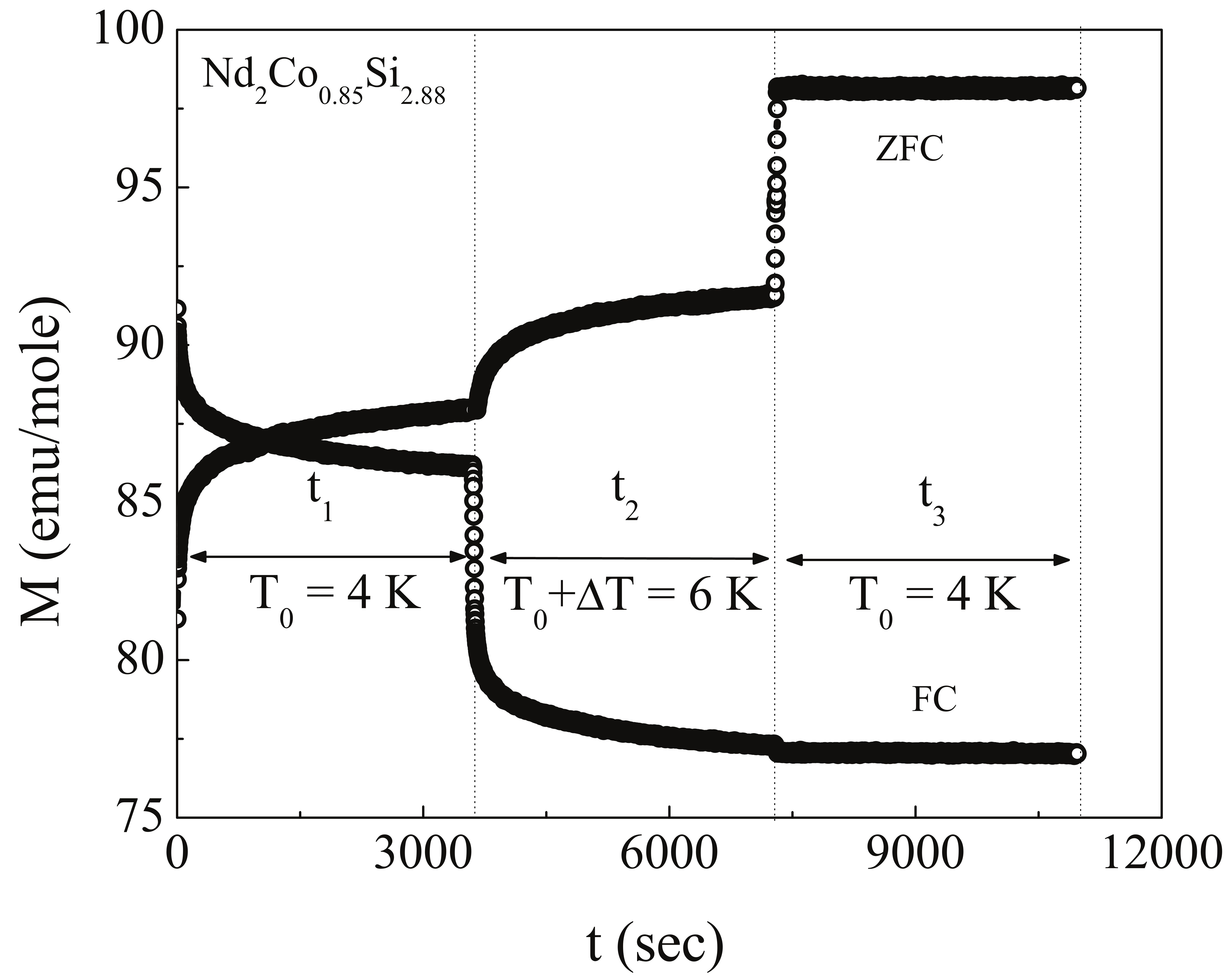}
\caption{Magnetic relaxation behavior of Nd$_2$Co$_{0.85}$Si$_{2.88}$ at 4 K with temporary heating at 6 K measured using ZFC and FC protocol for $H$ = 100 Oe.}\label{Nd_tempheating}
\end{figure}

The system behaves differently when it is temporarily heated to $T_0$ + $\Delta T$. In ZFC process, the sample was cooled from paramagnetic region to $T_{0}$ ($T_0 < T_{\rm L}$) in zero-field followed by $M(t)$ measurement during $t_1 = 1$~h at $T_0$ in presence of field. After lapse of $t_1$, the sample was heated to $T_0$ + $\Delta T$ with same applied field and $M(t)$ was recorded for time period $t_2 = 1$h. Finally, the temperature of the system was turned back to $T_{0}$ and $M(t)$ was measured for another time period $t_3 = 1$~h. In FC process the sample was cooled to $T_{0}$ in presence of magnetic field followed by $M(t)$ measurement during $t_1 = 1$~h after switching off the magnetic field. Then the sample was temporarily heated to $T_0$ + $\Delta T$ and $M(t)$ was recorded for $t_2 = 1$~h. The sample was finally cooled back to $T_{0}$ and $M(t)$ was measured for another $t_3 = 1$~h. From the Fig.~\ref{Nd_tempheating}, it is clear that both in the ZFC and FC protocol, magnetic relaxation behavior during $t_3$ is quite different from that during $t_1$. Thus the system can not restore its previous history during temporary heating implying absence of memory in the positive temperature cycle. Such asymmetric behavior of magnetic relaxation upon cooling and heating favours the hierarchical model of memory effect for the studied glassy compound Nd$_2$Co$_{0.85}$Si$_{2.88}$.

\section{Concluding Remarks}

Here, we report the successful synthesis of a novel triangular lattice compound Nd$_2$Co$_{0.85}$Si$_{2.88}$, that is stabilized in single-phase only with defect structure with vacancies. Magnetization measurements suggest the presence of two magnetic phase transitions in the system, at $T_H \sim 140$~K and at $T_{\rm L} \sim 6.5$~K. From our detailed studies on temperature, magnetic field, and time dependent magnetization, heat capacity, together with DFT calculations, we have argued that the high-temperature phase transition is associated with the development of ferromagnetic interaction among the itinerant Co-3$d$ moments, while the low-temperature transition is due to an antiferromagnetic coupling between localized Nd-4$f$ spin and itinerant Co-3$d$ spin sublattices. The long-range RKKY exchange interaction among the localized Nd-ions breaks due to the blocking of mediator $d$ electrons as conduction carriers, resulting in strong spin frustration in the compound. Spin fluctuations associated with the localized Nd-4$f$ ions start to develop below 30 K and compete with the itinerant FM interactions of Co spin-clusters, giving rise to an anomalous temperature dependence of magnetic coercivity. The detailed studies on low-temperature non-equilibrium dynamical behavior (including magnetic relaxation and associated aging phenomena along with magnetic memory effect both in ZFC and FC conditions) clearly suggest a short-range glassy state formation below $T_{\rm L}$ for the compound. The glassy component is originated due to the simultaneous presence of bond-disorder in crystal structure and strong spin frustration in the system. The unique mechanism of magnetic frustration discussed for Nd$_2$Co$_{0.85}$Si$_{2.88}$ is expected to contribute significantly to understand the origin and the mechanism of competing magnetic interactions as well as related phenomena in isostructural materials.

\section{Acknowledgements}

A major part of this work has been carried out and supported through CMPID project at SINP and funded by Department of Atomic Energy (DAE), Govt. of India. The authors are thankful to Anish Karmahapatra for technical support during XRD measurements. We are grateful to Shibasis Chatterjee and Tridib Das for SEM \& EDX measurements. The research at Ames Laboratory (in part) was supported by the U.S. Department of Energy, Office of Basic Energy Sciences, Division of Materials Sciences and Engineering. Ames Laboratory is operated for the U.S. Department of Energy by Iowa State University under Contract No.~DE-AC02-07CH11358. S.C. acknowledges University Grants Commission (UGC) for research fellowship. N. L. acknowledges the Council of Scientific and Industrial Research (CSIR), New Delhi for PPMS facility through XII Five Year Plan project MULTIFUN (CSC-0101) and thankful to the Director, CSIR-CECRI for the support (CECRI/PESVC/Pubs./2021-153).

\end{document}